\documentclass[prd,twocolumn,letterpaper,superscriptaddress]{revtex4}
\include{epsf}
\usepackage{graphicx}
\usepackage{pslatex}
\usepackage{url}

\def\deg{$^\circ$}

\begin{document}

\title{Design and Initial Performance of the Askaryan Radio Array \\
Prototype EeV Neutrino Detector at the South Pole}

\author{P.~Allison}
\affiliation{Dept. of Physics, Ohio State Univ., Columbus, OH 43210, USA.} 
\author{J.~Auffenberg}
\affiliation{IceCube Research Center, Univ. of Wisconsin- Madison WI 53703.}
\author{R.~Bard}
\affiliation{Dept. of Physics, Univ. of Maryland, College Park, MD, USA. }
\author{J.~J.~Beatty}
\affiliation{Dept. of Physics, Ohio State Univ., Columbus, OH 43210, USA.} 
\author{D.~Z.~Besson}
\affiliation{Dept. of Physics and Astronomy, Univ. of Kansas, Lawrence, KS 66045, USA. }
\author{S.~B\"oser}
\affiliation{Dept. of Physics, Univ. of Bonn, Bonn, Germany.}
\author{C.~Chen}
\author{P.~Chen}
\affiliation{Dept. of Physics, Grad. Inst. of Astrophys.,\& Leung Center for 
 Cosmology and Particle Astrophysics, National Taiwan Univ., Taipei, Taiwan.}
\author{A.~Connolly}
\affiliation{Dept. of Physics, Ohio State Univ., Columbus, OH 43210, USA.} 
\author{J.~Davies}
\affiliation{Dept. of Physics and Astronomy, Univ. College London, London, United Kingdom.}
\author{M.~DuVernois}
\affiliation{IceCube Research Center, Univ. of Wisconsin- Madison WI 53703.}
\author{B.~Fox}
\author{P.~W.~Gorham}
\affiliation{Dept. of Physics and Astronomy, Univ. of Hawaii, Manoa, HI 96822, USA. }
\author{E.~W.~Grashorn}
\affiliation{Dept. of Physics, Ohio State Univ., Columbus, OH 43210, USA.} 
\author{K.~Hanson}
\affiliation{Dept. of Physics, Univ. Libre de Bruxelles, Belgium.}
\author{J.~Haugen}
\affiliation{IceCube Research Center, Univ. of Wisconsin- Madison WI 53703.}
\author{K.~Helbing}
\affiliation{Dept. of Physics, Univ. of Wuppertal, Wuppertal, Germany.} 
\author{B.~Hill}
\affiliation{Dept. of Physics and Astronomy, Univ. of Hawaii, Manoa, HI 96822, USA. }
\author{K.~D.~Hoffman}
\affiliation{Dept. of Physics, Univ. of Maryland, College Park, MD, USA. }
\author{M.~Huang}
\affiliation{Dept. of Physics, Grad. Inst. of Astrophys.,\& Leung Center for 
 Cosmology and Particle Astrophysics, National Taiwan Univ., Taipei, Taiwan.}
\author{M.~H.~A.~Huang}
\affiliation{Dept. of Physics, Grad. Inst. of Astrophys.,\& Leung Center for 
 Cosmology and Particle Astrophysics, National Taiwan Univ., Taipei, Taiwan.}
\author{A.~Ishihara}
\affiliation{Dept. of Physics, Chiba university, Tokyo, Japan.}
\author{A.~Karle}
\affiliation{Dept. of Physics, Univ. of Wisconsin Madison, Madison, WI, USA.}
\author{D.~Kennedy}
\affiliation{Dept. of Physics and Astronomy, Univ. of Kansas, Lawrence, KS 66045, USA. }
\author{H.~Landsman}
\author{A.~Laundrie}
\affiliation{IceCube Research Center, Univ. of Wisconsin- Madison WI 53703.}
\author{T.~C.~Liu}
\affiliation{Dept. of Physics, Grad. Inst. of Astrophys.,\& Leung Center for 
 Cosmology and Particle Astrophysics, National Taiwan Univ., Taipei, Taiwan.}
\author{L.~Macchiarulo}
\affiliation{Dept. of Physics and Astronomy, Univ. of Hawaii, Manoa, HI 96822, USA. }
\author{K.~Mase}
\affiliation{Dept. of Physics, Chiba university, Tokyo, Japan.}
\author{T.~Meures}
\affiliation{Dept. of Physics, Univ. Libre de Bruxelles, Belgium.}
\author{R.~Meyhandan}
\author{C.~Miki}
\author{R.~Morse}
\affiliation{Dept. of Physics and Astronomy, Univ. of Hawaii, Manoa, HI 96822, USA. }
\author{M.~Newcomb}
\affiliation{IceCube Research Center, Univ. of Wisconsin- Madison WI 53703.}
\author{R.~J.~Nichol}
\affiliation{Dept. of Physics and Astronomy, Univ. College London, London, United Kingdom.}
\author{K.~Ratzlaff}
\affiliation{Instrumentation Design Laboratory, Univ. of Kansas, Lawrence, KS 66045, USA.}
\author{M.~Richman}
\affiliation{Dept. of Physics, Univ. of Maryland, College Park, MD, USA. }
\author{L.~Ritter}
\affiliation{Dept. of Physics and Astronomy, Univ. of Hawaii, Manoa, HI 96822, USA. }
\author{B.~Rotter}
\affiliation{Dept. of Physics and Astronomy, Univ. of Hawaii, Manoa, HI 96822, USA. }
\author{P.~Sandstrom}
\affiliation{IceCube Research Center, Univ. of Wisconsin- Madison WI 53703.}
\author{D.~Seckel}
\affiliation{Dept. of Physics, Univ. of Delaware, Newark, DE 19716.} 
\author{J.~Touart}
\affiliation{Dept. of Physics, Univ. of Maryland, College Park, MD, USA. }
\author{G.~S.~Varner}
\affiliation{Dept. of Physics and Astronomy, Univ. of Hawaii, Manoa, HI 96822, USA. }
\author{M.~-Z.~Wang}
\affiliation{Dept. of Physics, Grad. Inst. of Astrophys.,\& Leung Center for 
 Cosmology and Particle Astrophysics, National Taiwan Univ., Taipei, Taiwan.}
\author{C.~Weaver}
\affiliation{Dept. of Physics, Univ. of Wisconsin Madison, Madison, WI, USA.}
\author{A.~Wendorff}
\affiliation{Dept. of Physics and Astronomy, Univ. of Kansas, Lawrence, KS 66045, USA. }
\author{S.~Yoshida}
\affiliation{Dept. of Physics, Chiba university, Tokyo, Japan.}
\author{R.~Young}
\affiliation{Instrumentation Design Laboratory, Univ. of Kansas, Lawrence, KS 66045, USA.}
\collaboration{ ARA Collaboration}

\vspace{2mm}
\noindent


\begin{abstract}
We report on studies of the viability and sensitivity of the Askaryan
Radio Array (ARA), a new initiative to develop a Teraton-scale ultra-high energy
neutrino detector in deep, radio-transparent ice near Amundsen-Scott station
at the South Pole. An initial prototype ARA detector system was installed
in January 2011, 
and has been operating continuously since then. We report on studies
of the background radio noise levels, the radio clarity of the ice, and the
estimated sensitivity of the planned ARA array given these results, based 
on the first five months of operation. 
Anthropogenic radio interference in the vicinity of the South Pole currently
leads to a few-percent loss of data, but no overall effect on the background
noise levels, which are dominated by the thermal noise floor of the cold
polar ice, and galactic noise at lower frequencies. 
We have also successfully detected signals originating
from a 2.5~km deep impulse generator at a distance of over 3~km from our
prototype detector, confirming prior estimates of kilometer-scale 
attenuation lengths for cold polar ice. These are also the first such measurements
for propagation over such large slant distances in ice. Based on
these data, ARA-37, the 200~km$^2$ array now under construction, will achieve the
highest sensitivity of any planned or existing neutrino detector in the
$10^{16}-10^{19}$~eV energy range.
\end{abstract}
\pacs{95.55.Vj, 98.70.Sa}
\narrowtext 

\maketitle

\section{Introduction}

At PeV energies and above (1~PeV = $10^{15}$~eV), 
ultra-high energy (UHE) neutrinos can be most
efficiently detected in dense, radio-frequency 
(RF) transparent media via the Askaryan effect~\cite{Ask62}.
The abundant cold ice covering the
geographic South Pole, with its exceptional RF clarity, has been host to several 
pioneering efforts to develop this approach, including RICE~\cite{RICE06} and 
ANITA~\cite{ANITA2,ANITAinst}, which trace their roots back to 
suggestions made in the mid-1980's~\cite{Gusev}.
Within the last year, a multi-phased experiment
(the Askaryan Radio Array, or ``ARA'')  has been initiated at the South Pole~\cite{whitepaper}, 
designed to ultimately accumulate hundreds of {\it GZK neutrinos}: 
those neutrinos which arise from the cumulative interactions of
ultra-high energy cosmic rays (UHECRs) throughout the universe,
specifically through the Greisen-Zatsepin-Kuzmin, or GZK process~\cite{Greisen,ZK}.
Combining the drilling and logistics experience of IceCube, 
the {\it in situ}  RF experience of RICE and the sub-nanosecond-scale data 
acquisition and triggering of ANITA, the singular goal of 
Phase 1 of ARA is to make the first definitive observation
of the cosmogenic GZK neutrinos with a radio antenna array covering a roughly
200~km$^2$ area, denoted here as ARA-37. 

In the longer-term, the modularity of the ARA design will 
engender cost-effective growth of the array to the Teraton scale necessary
to carry out an observatory-class UHE neutrino science program with 
broad intellectual merit, comprising basic 
particle physics, including neutrino flavor and cross-section studies in
previously unprobed kinematic regimes; 
astrophysics such as tests of models of galactic evolution, as well
as cosmic acceleration and propagation 
models relevant to ultra-high-energy cosmic ray (UHECR)
observatories such as the Pierre Auger and Telescope Array Experiments.

A critical question for any radio-based instrument deployed
at the South Pole is whether the local man-made radio interference
will prevent such an instrument from reaching its science goals. In
addition, {\it in situ} measurements of the radio noise floor, and
the radio transmissivity of the ice, as well as its propagation
characteristics, are necessary to assess the ultimate sensitivity
of ARA. 

To address these issues, we have deployed an initial 
ARA detector station in prototype form, which we also denote the ARA Testbed,
with much of the functionality planned for the ARA-37 design. 
One of the features of the current
ARA design is that each element of the large array comprises
a standalone neutrino detector for its surrounding ice. Although
multiple-station events are in no way excluded, they are not
required for either triggering or reconstruction of the 
interaction vertex or neutrino direction. This is done to
maximize the sensitivity of the array at the expense of some
angular and energy resolution. Once UHE neutrino fluxes and
first-order energy spectra are established with a discovery
instrument, then future installations can be further optimized
for resolution as well as sensitivity.

In our
current design, a single array element, denoted as a {\it station}
here, consists of a cluster with of order 16 embedded antennas,
deployed up to 200~m deep in several vertical boreholes placed with 
tens-of-meter horizontal spacing to form a small sub-array.
We report here on the
initial deployment of our first station, and results of 
measurements over the first four months of its operation.
In all aspects the environment and ice quality have exceeded
our expectations, and the prospects for ARA-37 thus appear 
excellent.

\section{Instrument Design}

The ARA stations must be designed to detect a highly linearly-polarized RF impulse
arriving from any direction, either as an approximate plane wave (for distant
sources) or a spherical wave for sources that are closer. The individual antennas
that comprise a station must have response that is preferably dipole-like in
angular extent, but the antenna array must also be sensitive in two orthogonal
polarizations to avoid any bias against detecting arbitrary planes of polarization,
since the RF impulses that arise from the Askaryan effect may be observed
at an arbitrary angle with respect to the plane of polarization. 

Because of 
practical limitations on drilling, the antennas should fit within a 15~cm
diameter borehole, and they should be at a depth below the firn,
the top layer of the ice sheet made of compacted snow of density lower than that
of solid ice, and extending down typically 150~m at the South Pole. 
Antennas within the firn are subject to ray-bending within the 
refractive index gradient which
complicates the triggering and reconstruction, and also leads to an effective
inverted horizon, cutting off the detectability of more distance sources
in the ice. Since South Polar ice has its highest RF clarity in the coldest
ice, the antenna array response should be optimized for detection in the
upper 1 to 1.5 km of ice, although deeper events will also be seen. Since drilling
costs increase dramatically with the required depth of the hole, there is
clear cost-benefit in making the holes no deeper than needed to achieve
the science goals. The array must also guard carefully against any self-noise
radio interference, that is: all data acquisition, power, and transmission
lines must be electromagnetically shielded with extreme care. 

\begin{table*}[hbt!]
\caption{Design specifications for ARA, and the implementation for the prototype station.
\label{design1}}
\vspace{3mm}
 \begin{footnotesize}
  \begin{tabular}{lcr}
\hline \hline
{ {\bf Specified parameter}} &   {\bf ARA 2012++ planned}    & {\bf ARA 2010-2011 prototype} \\ \hline
Number of Vpol antennas &  8 &  2 near-surface, 4 in ice  \\
Vpol antenna type & bicone & bicone \\
Vpol antenna bandwidth (MHz) &  150-850 & 150-850 \\
Number of Hpol antennas & 8 &  2 near-surface, 6 in ice \\
Hpol antenna type & quad-slotted cylinder & bowtie-slotted-cylinder \\
Hpol antenna bandwidth (MHz) &  200-850 & 250-850 \\
Number of Surface antennas & 4 & 2 \\
Surface antenna type & fat dipole & fat dipole \\
Surface antenna bandwidth (MHz) &  30-300 & 30-300 \\
Number of signal boreholes &  4  & 6 \\
Borehole depth (m) & 200 & 30 \\
Vertical antenna configuration & H,V above H,V & V or H above H or V \\
Vertical antenna pair spacing (m) & 20 & 5 \\
Approximate geometry & trapezoidal & trapezoidal \\ 
Approximate radius (m) & 10 & 10 \\
Number of calibration antenna boreholes & 3 & 3 \\
Calibration borehole distance from center (m) &  40 (2), 750 (1) & 30 \\
Calibration borehole geometry & isosceles triangle & equilateral triangle \\
Calibration signal types & noise and impulse & impulse only \\
LNA noise figure (K) &  $< 80$ & $<80$\\
LNA/amplifier dynamic range & 30:1 & 30:1\\
RF amplifier total gain (dB) & $>75$ & $>75$ \\ \hline
  \end{tabular}
 \end{footnotesize}
\end{table*}

In Table I, we list the primary instrument specifications chosen for the ARA station design,
giving key parameters such as the geometry and antenna requirements.
The ARA prototype instrument shares much of the functionality of the planned complete
ARA stations, with several notable differences, which are enumerated in the Table.

\subsection{Overall system}

\begin{figure*}[ht!]
 \begin{center}
\includegraphics[width=5.in]{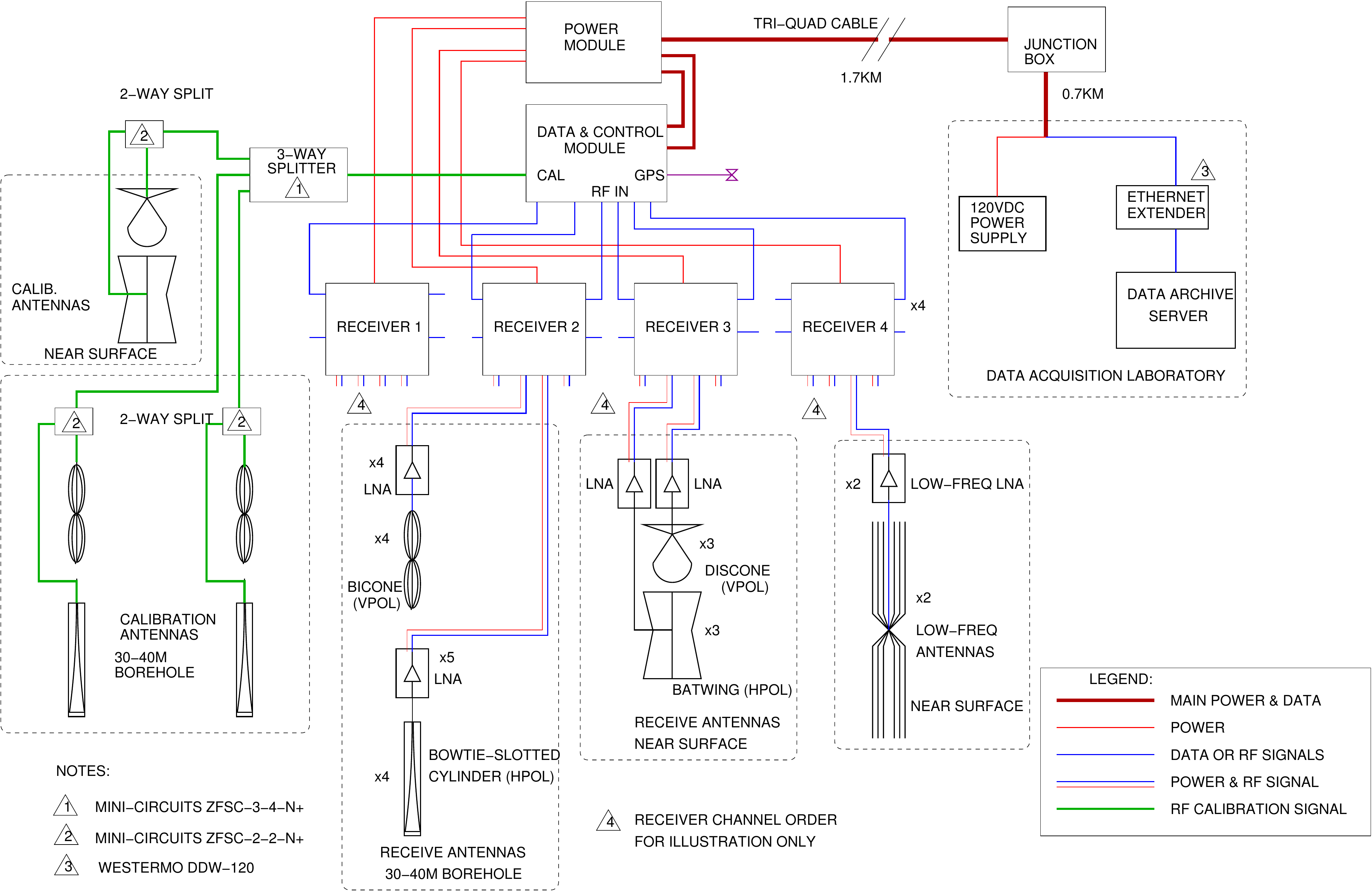}
 \caption{Block diagram of the entire ARA prototype system.}
 \label{ARAblock1}
 \end{center}
 \end{figure*}

A block diagram of the installed ARA testbed station is shown in Fig.~\ref{ARAblock1}. 
A variety of different antennas, both downhole and near-surface, are used. In each
case a low-noise preamplifier is inserted in close proximity to the antenna
to minimize insertion loss and associated thermal noise. Highly-shielded coaxial cables
(Andrew Heliax or Times Microwave LMR-series cables) then lead the signals to a secondary
receiver which boosts the signal with another 40~dB amplifier to bring the thermal noise
levels (which are several $\mu$V referenced to the LNA inputs) to several tens of mV RMS
at the DAQ system. The signal is split for separate trigger and waveform recording paths.
For each trigger, the RF waveforms and associated housekeeping (temperatures, threshold 
settings, antenna singles rates, etc.) are recorded locally by a single-board computer,
then transmitted over twisted pair wires via an ethernet modem to a data receiving
computer at the South Pole station. There the data is repackaged for transmission
via satellite link to northern hemisphere computer archives. Future installations will use
an optical fiber or wireless data link for the data transmission to and from the individual stations.

    \subsection{Antennas}

\begin{figure*}[ht!]
 \begin{center}
\includegraphics[width=6.5in]{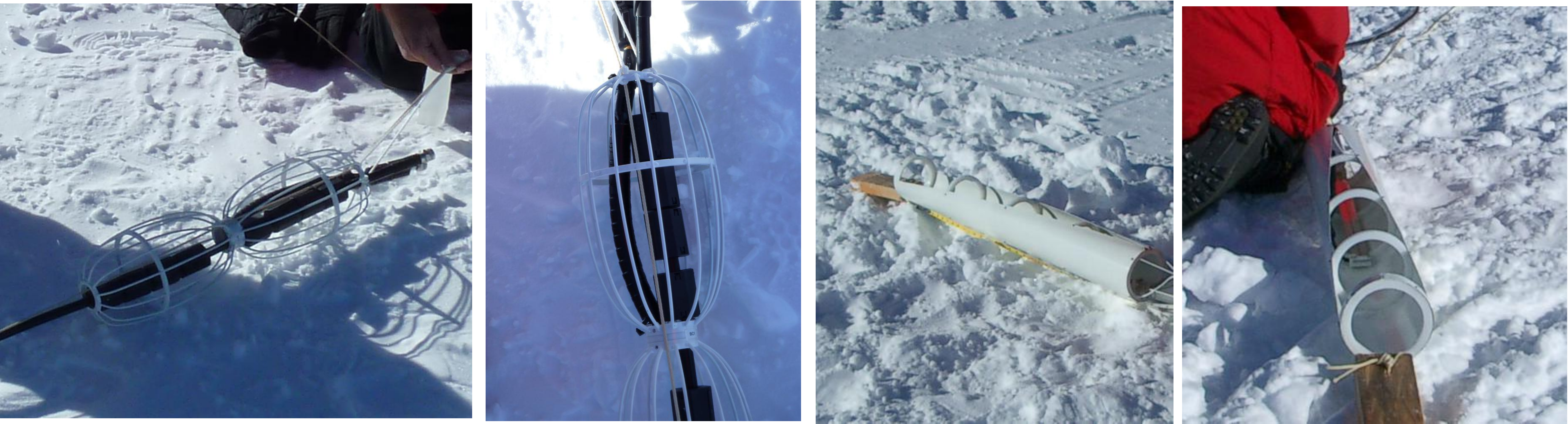}
 \caption{ARA testbed downhole antennas: left two images, wire-frame bicone Vpol antennas;
right two images, bowtie-slotted-cylinder Hpol antennas.}
 \label{antennas}
 \end{center}
 \end{figure*}

 \begin{figure*}[ht!]
 \begin{center}
\includegraphics[width=6.5in]{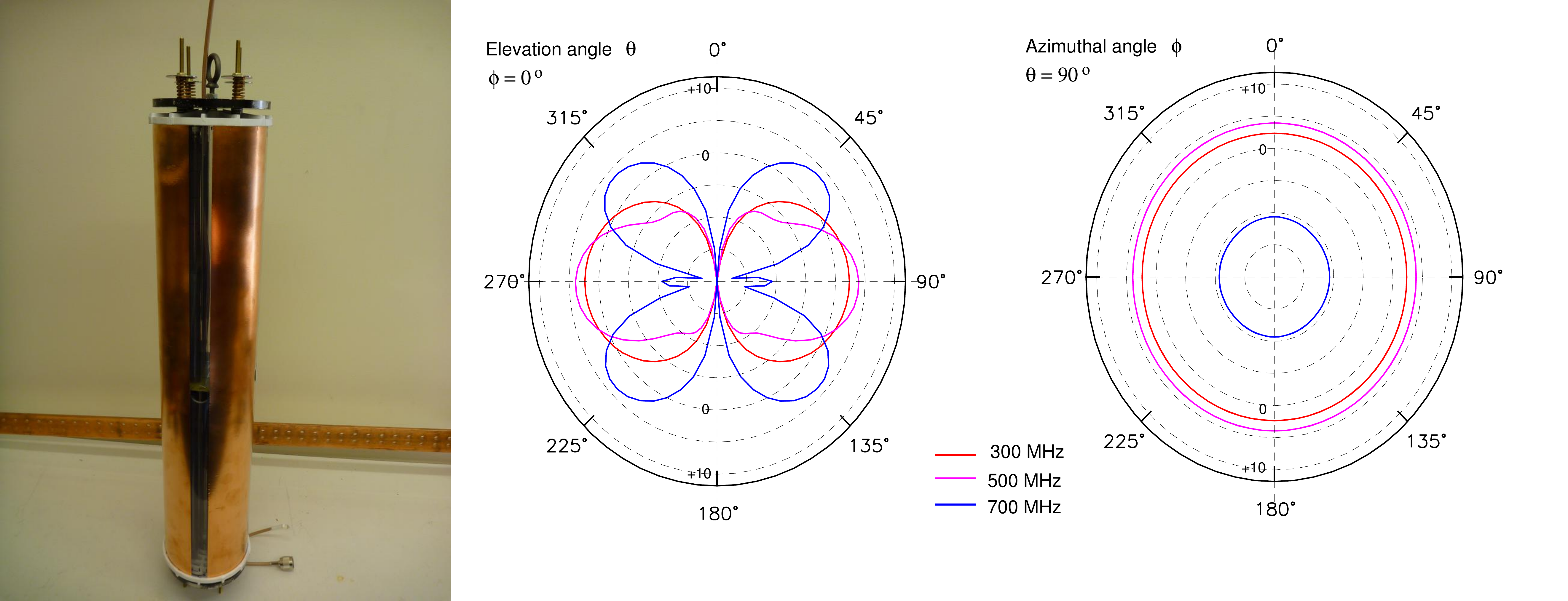}
 \caption{Left: Quad-slot cylinder antenna used in one borehole for ARA-testbed.
Center: Simulated Gain (dBi) vs. elevation angle ( zero degrees is the vertical direction) for three frequencies for the QSC antenna. 
Right: Simulated Gain (dBi) in the horizontal plane vs. azimuth, showing the high degree of uniformity of the
QSC azimuthal response.}
 \label{QSC}
 \end{center}
 \end{figure*}

Antenna designs for ARA are highly constrained by the need to fit into a borehole in the ice.
Currently, ARA has demonstrated the ability to successfully drill 15~cm diameter boreholes
of the required 200~m depth. This hot-water drilling operation requires pumping
out of the drill water after completion of the hole, to avoid freeze-in of ARA
equipment, at least in the early phases of array construction where recovery of
downhole instrumentation when necessary helps to improve reliability. For our initial
season, equipment for pumping the holes dry was not available yet, and thus the initial
season's boreholes were drilled only to a depth of order 30~m typically, where pooling
of water prevented deeper drilling. However, the 15~cm diameter for these holes was
maintained.

 \begin{figure}[ht!]
 \begin{center}
\includegraphics[width=3.3in]{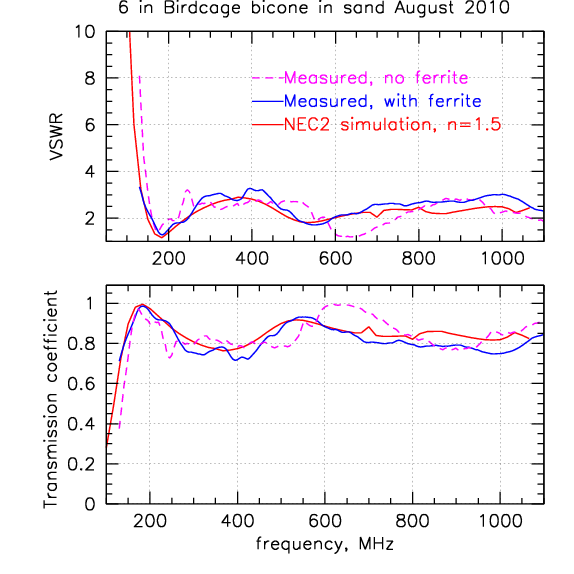}
 \caption{Measured and predicted Voltage standing wave ratio (VSWR, top) 
and equivalent power transmissivity (bottom)
of the bicone antenna, with and without the internal ferrite loading on the
feedline. The ferrite loading is necessary to prevent coupling of
RFI via stray induced currents the external shield; these results show that the presence
of the ferrites within the antenna did not degrade the performance.}
 \label{Birdcage_eff}
 \end{center}
 \end{figure}

 \begin{figure}[ht!]
 \begin{center}
\includegraphics[width=3.3in]{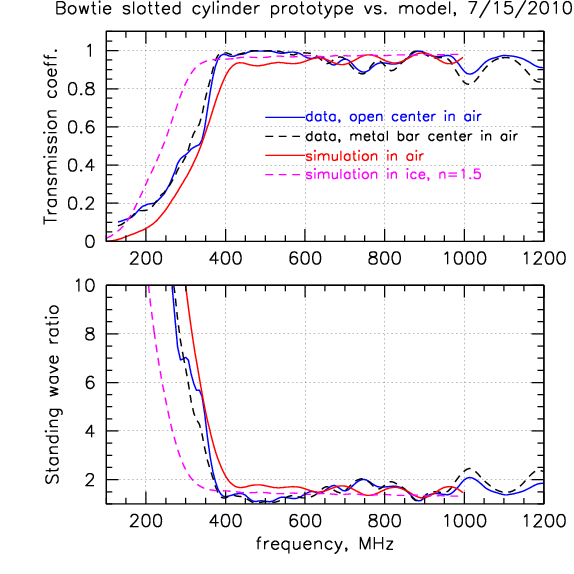}
 \caption{Measured and predicted Voltage standing wave ratio (VSWR, top) 
and equivalent power transmissivity (bottom)
of the bowtie-slotted-cylinder antenna, along with simulations in air and
in ice. Measurements were also made with a metal bar passing through
the center of the antenna to confirm its immunity to an internal conductor.}
 \label{BSC_eff}
 \end{center}
 \end{figure}

An additional issue for borehole antennas where there are multiple antennas along
the hole, fed by cables from the surface, is that the conductive coaxial cables
must pass through upper antennas to reach lower ones. This constraint forces the
need for a central hole in the antenna to ensure a uniform azimuthal response, 
and these holes require re-design of the
antenna feed region for many traditional borehole antenna designs.
With the diameter constraint, antennas for vertical polarization are straightforward:
a fat-dipole or related biconical designs will provide very broadband, well-matched 
antennas. With the additional cable-passthrough constraint, it is more difficult to
find a design that retains azimuthal symmetry while still passing the signal cable,
but we have settled on a hollow-center biconical design, where the feed region is
annular around the passthrough cables, and is fed at multiple locations along the
annulus, with appropriate impedance matching. 

For a corresponding antenna with horizontal polarization (Hpol) two designs were
implemented for testing in the ARA testbed: a bowtie-slotted-cylinder (BSC) antenna,
and a quad-slotted-cylinder (QSC) antenna with internal ferrite loading to effectively lower
its frequency response.  The goal for both sets of antennas was to cover
a frequency range from about 150~MHz to 850~MHz. This goal was achieved with the
Vpol antennas, but the 15~cm diameter borehole constraint has proved challenging
for the Hpol antennas, both of which have difficulty getting frequency response
below about 200-250~MHz in ice. In addition, the BSC antenna, although it was found
to have better efficiency than the QSC, suffers from some azimuthal asymmetry in its
response, and thus the QSC, which has uniform azimuthal response, will be used
for future ARA stations. In the current testbed station, we have primarily used the
BSC antennas because of the ease of their manufacture for the 2011 season.
Figure~\ref{antennas} shows photographs of the wire-frame bicone antennas and the
BSCs as they were readied for deployment. Fig~\ref{QSC} shows a photo of one of
the QSC prototypes (only one of the 4 slots is evident), along with simulated results
for the gain patterns in elevation and azimuth, illustrating the uniformity, which
was confirmed at several angles in laboratory measurements.

Figures ~\ref{Birdcage_eff} and \ref{BSC_eff} show the voltage standing wave ratio (VSWR),
along with the power transmission coefficient for the primary borehole antennas used
for the ARA-testbed. VSWR is related to the complex voltage reflection coefficient $\rho$ of the
antenna via the relation
$$VSWR(\nu) = \frac{|\rho(\nu) + 1|}{|\rho(\nu) - 1|}$$
and the effective power transmission coefficient $T$ (either as a receiver or transmitter from
antenna duality) is given by
$$ T(\nu) = |1 - \rho(\nu)|^2$$
and may be thought of as the effective quantum efficiency of the antenna vs. frequency $\nu$
although RF antennas in the VHF to UHF range never operate in a photon-noise limited 
regime.

 \begin{figure}[ht!]
 \begin{center}
\includegraphics[width=3.3in]{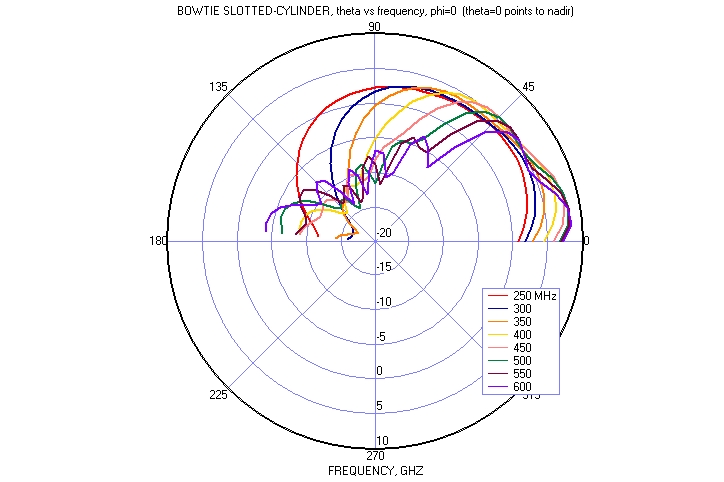}
\vspace{2mm}
\includegraphics[width=3.3in]{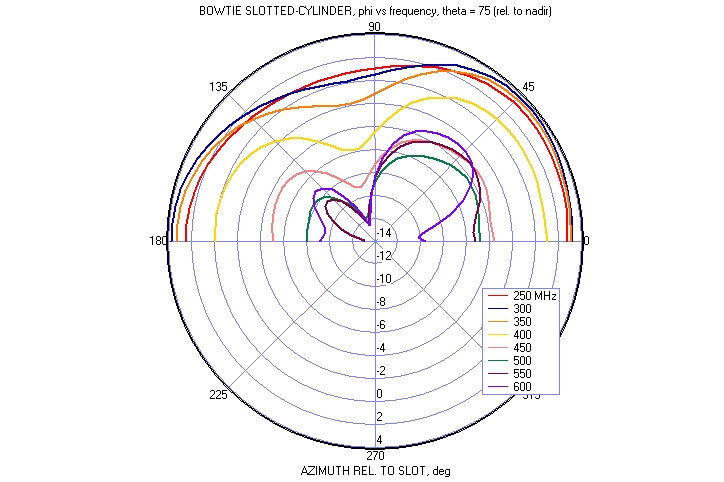}
 \caption{Top: BSC gain for a range of frequencies, as a function
of polar angle. The feed point of the antenna is to the left, 
and the antenna is aligned with the $x$-axis here. Bottom,
azimuthal response for a similar range of frequencies.}
 \label{BSC_gain}
 \end{center}
 \end{figure}

In addition to the coupling efficiency of the antennas, the other important parameter for
RF performance is the antenna directivity gain $G$, often denoted as just gain, and related
to the effective power collection area of the antenna via the fundamental relation
$$A_{eff}(\nu) = \frac{G c^2}{4\pi\nu^2}$$
for the speed of light $c$. $A_{eff}$ is in turn related to the vector effective height of 
the antenna by
$$ \vec{h}_{eff} = 2~\sqrt{\frac{R_{r}~ A_{eff}}{\eta}}~~\hat{h} $$
where $R_r$ is the radiation resistance of the antenna, $\hat{h}$
is a unit vector representing the antenna axis, and $\eta = \sqrt{\mu/\epsilon}$
is the impedance of the medium, $\eta \approx 120\pi~\Omega$ in free space~\cite{Kraus}.
For simple antennas such as dipoles, the induced voltage into a matched receiver load
is given by $V = \vec{E} \cdot \vec{h}_{eff} / 2$, and for our borehole antennas
$\hat{h} = \hat{z}$, where $\hat{z}$ is the local vertical direction.
For our horizontally-polarized antennas which rely on coupling to a slot (the Babinet dual
to a dipole), the coupling may either be thought of as a magnetic induction to the
slot, or as electric coupling to an effective height which is toroidal, {\it e.g.}
$\hat{h} = \hat{\phi}$ where $\hat{\phi}$ is the azimuth unit vector in cylindrical 
coordinates. In either case, for our broadband antennas which have more
complex frequency-dependence, detailed estimates of the antenna response wil require an integral 
over the frequency-dependent behavior.
For antennas immersed in a dielectric, both the speed of light and impedance are
affected. The dielectric generally lowers the frequency response of the antenna
compared to air.

For our biconical antennas, we found the gain to be commensurate with that expected
from a theoretical bicone, with a primary mode that behaves as $\cos^2$ of the polar angle,
and extends for about 1 octave in frequency compared to the turn-on of the antenna.
At higher frequencies, the response becomes multi-modal as expected from a dipole used
at its higher-order resonance regions, but the frequency bandwidth of these modes is
much larger than for a resonant dipole, due to the broadband biconical response.
The QSC Hpol antenna gain behaves in a very similar manner to the bicones, except that
its turn-on frequency is higher; this is consistent with the slot antennas performing
as a Babinet complementary antenna to a dipole.

For the BSC antenna, the response is more like an inverted broadband monopole, as
shown in Fig.~\ref{BSC_gain}. In the top pane, the polar angle response is shown.
The antenna is oriented horizontally for this plot, but in the borehole it is
placed vertically with the high-frequency response peaking down, giving the antenna
higher sensitivity to upcoming RF at high frequencies. In azimuth there is also 
a front-to-back asymmetry which leads to better response for incoming RF for
the side with the open slot.

 \begin{figure}[ht!]
 \begin{center}
\includegraphics[width=3.25in]{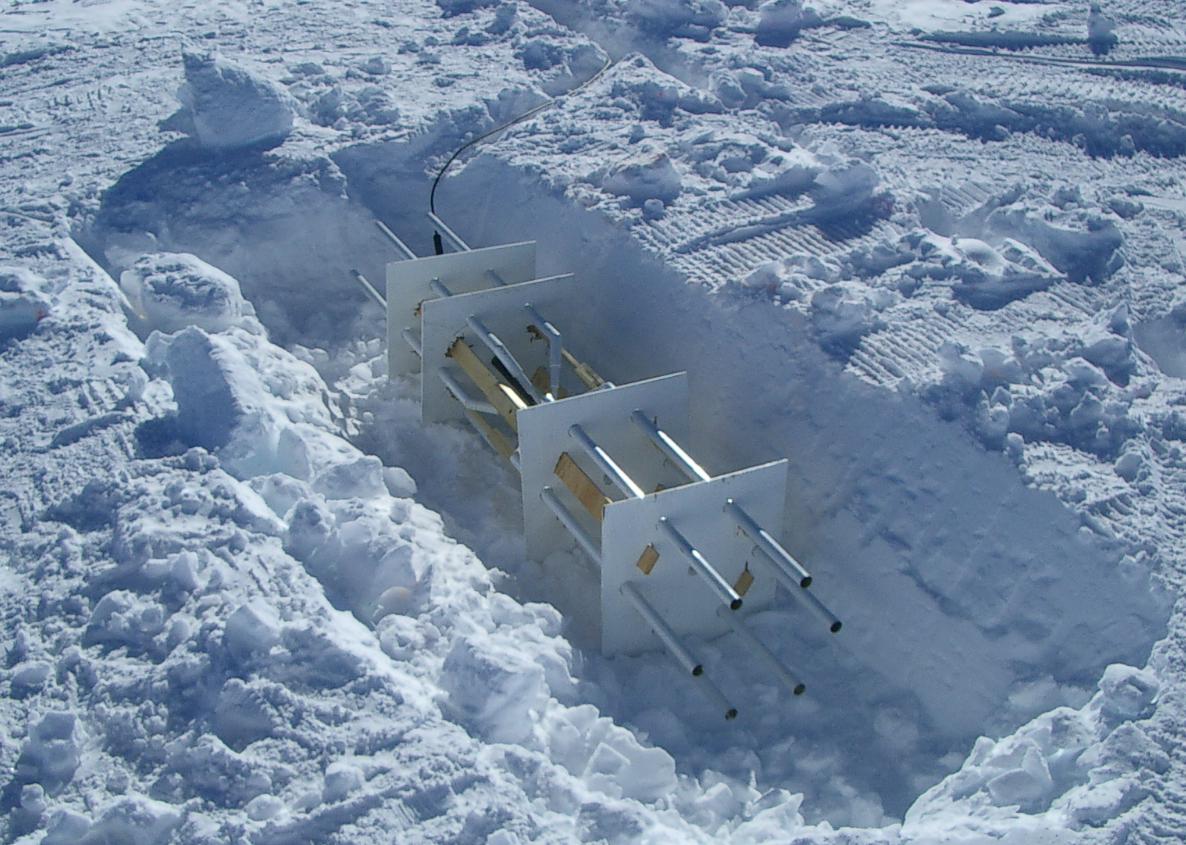}
 \caption{Fat dipole used for low-frequency measurements in
ARA-testbed. The antenna is about 0.5 m in diameter and 4m long, and
is installed just below the surface in the trench shown.}
 \label{deathray}
 \end{center}
 \end{figure}

Several other antennas were employed in the ARA testbed. Several larger discone (Vpol) and
batwing (Hpol) designs were submerged about 1~meter below the surface, and these performed 
well, but as they require a 40~cm diameter borehole they are impractical for future use.
We also deployed two surface fat-dipole (SFD) antennas (Fig.~\ref{deathray} 
with a frequency response extending over
a much lower range than our borehole antennas: 30-300~MHz. These were emplaced to 
assess the environment for detection of geosynchrotron RF emission from 
cosmic ray extensive air showers (EAS) which arrive from interactions in the atmosphere
above the array. Such RF impulses have predominantly lower frequency content than
Askaryan impulses~\cite{ANITA-UHECRs}, but have RF spectra which can extend to hundreds of MHz and thus
will be detected in the deeper antennas as well. The SFDs in this case can function
as a way to conclusively tag EAS radio events and distinguish them from other
RF interference, if the backgrounds are low enough in the VHF band.

    \subsection{Preamplifiers and receivers}

The ARA testbed preamplifier module consists of a front-end passband and notch filter
designed to block any out-of-band signals and to notch one strong in-band signal
from the South Pole communication system at about 450~MHz, and a low-noise amplifier (LNA). 
Any insertion loss  in the front-end filter causes a loss of signal combined with
an increase in thermal noise, and this filter was thus carefully designed to minimize
in-band losses to a fraction of 0.5~dB or less. The notch and the out-of-band rejection
of the filter are necessary to prevent unwanted anthropogenic noise from saturating the
LNA and driving it into a nonlinear operating range where its gain would be compressed.

The combination of the front-end filter and LNA contributes about 80K of 
input-referenced noise to the signal path, and the external thermal noise of
the ice at an ambient temperature of 230K (the average extends out to a volume of ice
of order 1 attenuation length in radius), giving a resulting system noise of order
310K expected. This leads to an input power of about -85~dBm for the LNA.
Once the signals are amplified by the LNA, they are transmitted to the top of the
borehole where a 2nd stage receiver amplifier with gain of about 40~dB boosts the
signals again to a range where they can be digitized.

 \begin{figure}[ht!]
 \begin{center}
\includegraphics[width=3.5in]{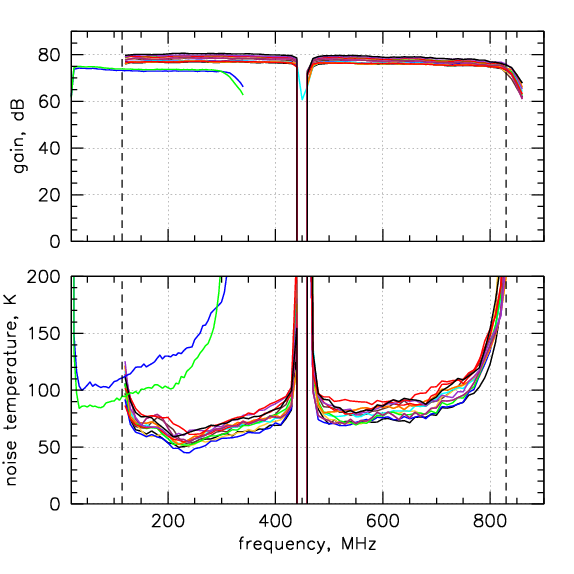}
 \caption{Total gain and noise figure for ARA prototype preamplifiers+receivers.
the two low-frequency antenna LNAs appear to the leftmost part of the plot.}
 \label{GNF}
 \end{center}
 \end{figure}

The gain and noise figure of the preamplifier+receiver signal chain was measured for
all ARA testbed channels using an Agilent HP8970A noise figure meter with an
NIST-traceable calibrated noise source. These measurements were done with
the devices under test at an ambient temperature of about -15C. This
is warmer than the expected temperatures in the ice and at the surface during the
austral winter, and thus separate measurements were made in a freezer at -55C to
determine the additional temperature scaling for the noise figure and gain.

The results for the noise figure and gain as a function of frequency 
are shown in Fig.~\ref{GNF}. The complete
system gain is close to 80~dB as expected, and the noise figure of the preamplifier
modules is in the range of 70-90K over most of the band, except for the
region of the notch at 450~MHz, and the high edge of the band where both
cable and low-pass filter attenuation begins to dominate. Given that the
ambient temperature of the ice is still the dominant contribution to the
overall system noise, these results are well within the specifications.

    \subsection{Trigger, Digitizer, and Data harvesting}
\begin{figure*}[ht!]
 \begin{center}
\includegraphics[width=6.5in]{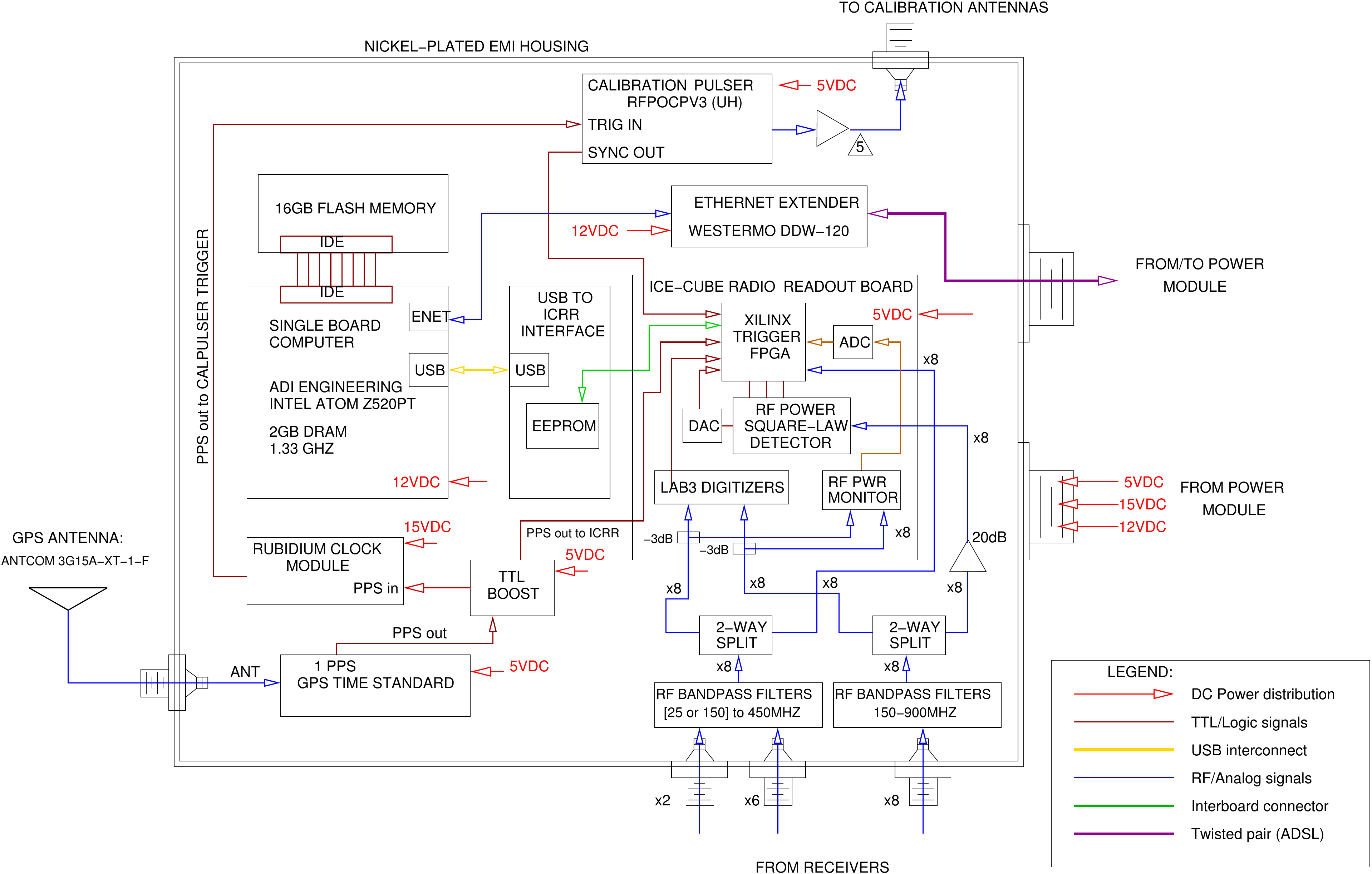}
 \caption{Block diagram of the ARA remote data acquisition system.}
 \label{ARAblock2}
 \end{center}
 \end{figure*}

Although it is currently possible to continuously digitize RF signals and
use these digital waveforms to assess the possible presence of impulsive
events of interest, the power requirements for such a streaming trigger
are still prohibitive for our design goals, which require that future  ARA
stations be ultimately able to function at remote locations far removed from
large-scale power infrastructure. Thus we only digitize waveforms when
a separate analog system determines that some interesting pattern of RF signals
has crossed above an intensity threshold. 

The ARA-testbed system uses coaxial tunnel diode detectors to provide
a unipolar signal proportional to the instantaneous RF power over a few nanosecond
time scale, and these pulses are fed into a discriminator implemented via
a field-programmable gate array (FPGA) using low-voltage differential signal
comparators. The FPGA then generates digital one-shot signals for each
of the 16 input antennas, and these digital outputs can in turn be
assessed via firmware logic to determine if a trigger condition is met.
For our current system, the primary trigger is set so that when 
any 3 of the 8 primary
Vpol and Hpol borehole antennas produce signals crossing threshold
within a 100~ns window, the trigger condition is satisfied, and
all waveforms are digitized and read out by the 
data acquisition system (DAQ) which is shown in block diagram 
form in Fig.~\ref{ARAblock2}.

\begin{figure}[ht!]
 \begin{center}
\includegraphics[width=3.5in]{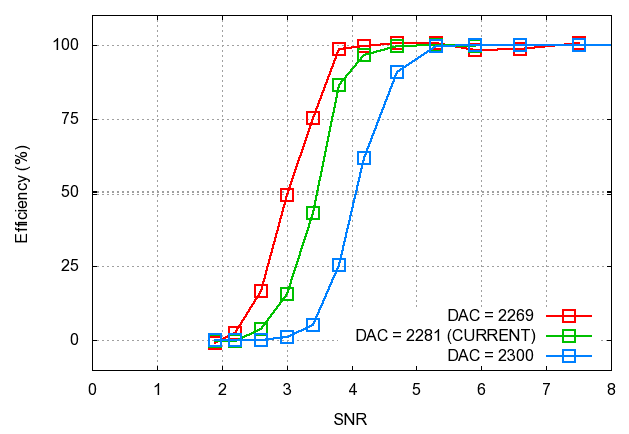}
 \caption{Overall trigger efficiency vs. the voltage signal-to-noise ratio of the
input RF impulse. The SNR is measured with respect to the RMS receiver voltage for
the baseline thermal noise level}
 \label{ARA-trigger}
 \end{center}
 \end{figure}

The trigger condition of any 3 of the primary 8 borehole antennas crossing
a pre-set threshold leads to a strong dependence of the overall global trigger
rate on the individual threshold-crossing rates, or singles-rates $R_{single}$,
of each antenna, which in turn is determined both by the average
power level in each antenna, and by the value and stability of the 
pre-set threshold. To first order the global trigger rate is given by:
\begin{equation}
R_{k:N} = \sum_{k=3}^N {N \choose k}~ R_{single}^k ~ \tau^{k-1}
\end{equation}
and $\tau$ is the coincidence window, 100~ns in our case, and $N=8$
is the number of antennas that contribute to the coincidence.
From this it is evident that the global rate depends on at least
the cube of the singles rate. Since we are triggering on fluctuations in
the antenna power which follow an exponential distribution, 
$R_{single} \propto \exp(-V_{thresh})$~\cite{Goodman}, 
and the global trigger rate
will depend on the cube of an exponential of the voltage threshold,
or the underlying power level, 
illustrating the degree of sensitivity to both thermal noise changes and
threshold stability. For this reason, it is generally advisable to implement
a {\it noise-riding} threshold, which servos (at a typical cadence below 1 Hz)
the voltage thresholds to
maintain approximately constant $R_{single}$, while recording the thresholds
used. This creates a nearly negligible change in overall sensitivity but
leads to more stable operation. Such a noise-riding system was
designed into our current trigger, but has not been yet implemented in
the operating period we describe here. In fact we do see evidence of
temperature-dependent drifts in global trigger rate in our current data,
as will be shown in a later section.

Our current DAQ system has a maximum sustained trigger rate of 25~Hz due to limitations
of the readout system that will be improved substantially for future stations.
However, rates above 10~Hz incur substantial deadtime for the system, so
we currently set thresholds to keep the rate below a few Hz. This trigger level
is quite sensitive, however, as shown by Fig.~\ref{ARA-trigger}. Here
the trigger efficiency is shown as a function of the SNR of the
received signal relative to the thermal noise level of the system, which
constitutes the ultimate noise floor. Our current system is 50\% efficient
for any incoming impulse which gives antenna voltage 
amplitude exceeding $3.5\sigma$ above
thermal noise.

12-bit waveform analog-to-digital conversion is performed using the
Large Analog Bandwidth Recorder
and Digitizer with Ordered Readout (LABRADOR), an Application Specific
Integrated Circuit (ASIC) developed at the Univ. of Hawaii~\cite{Varner}.
In our implementation of the LABRADOR, eight of the channels are sampled
at 2 Gigasamples/second effectively, with a 1~GHz Nyquist cutoff, and
the remaining eight channels are sampled at 1~Gsample/sec with a
500~MHz Nyquist cutoff. The latter channels are primarily the
surface and near-surface antennas. Near-future implementations of 
a new ASIC for ARA will extend the high-rate sampling and bandwidth to
all 16 channels.

Once digitized, the data is transferred from the LABRADOR to a
single-board computer and packetized.
Data is then transferred via an ethernet extender over 1.8~km of twisted pair cable
from the testbed DAQ computer to a server residing in
a South Pole laboratory, where it is archived on disk. A selection of data
is then packaged for transfer via a satellite link to northern hemisphere
data servers.

 \subsection{{\it In situ} Calibration}

To ensure that the triggering and data acquisition system are functioning
properly, the trigger threshold is set such that we trigger at a low
but quasi-continuous rate on thermal accidentals. This rate is typically
1~Hz or less. In addition, the system is software-triggered at 0.5~Hz to
provide a continuing record of unbiased waveform samples of the RF background. 
Finally, a GPS-synchronized Rubidium clock triggers a local calibration
pulser that sends $\sim 250$~ps impulse to antennas in a borehole about 30~m radius 
from the center of the testbed array, and this impulse is split and transmitted
from both Vpol and Hpol antennas to provide an impulse calibration for the
array. Arrival times for these impulses at each antenna help to determine
residual inter-antenna delays for the array, and this in turn is used to model
the local index of refraction of the ice.

    \subsubsection{Deep IceCube calibration pulser}

To date all results of long-path ice transparency measurements for the
Antarctic ice sheets have been made via
transmission paths that are largely vertical through reflections off the
basal interface at the bottom of the ice mass~\cite{icepaper}. Such measurements 
produce conservative lower limits to the average integrated ice attenuation length.
If the bottom reflection coefficient is known, the resulting attenuation
length measurement can be stated with higher confidence, although it still
represents an average over a large range of ice temperatures. For South Pole ice,
experiments in this vein have yielded several results on the ice 
attenuation that indicate km-scale attenuation lengths in the 
upper km of ice~\cite{icepaper,Ice08,Ice09,Ice2011}.

For ARA we are interested not just in near-vertical propagation, but propagation
over a full range of angles, since neutrino interactions can take place throughout
the ice target mass. To begin to address measurements that will support 
horizontal or slant propagation, we were able in the 2010-2011 season, with the
help and cooperation of the IceCube project, to deploy a set of pulse generators
and antennas at several locations near the ARA edge of the IceCube detector, at
ice depths of up to 2450~m. These pulser systems used several kV-peak impulses
transmitted through a vertically-polarized biconical antenna. Estimates of the
signal strength received in the ARA-testbed using conservative values for the
ice propagation were used to size the choice of transmitter amplitude, but with
some risk that, if the ice were much more transparent than expected, the
ARA-testbed might see saturation in the received signals. This risk was accepted
since stations further away from IceCube would still receive signals within
acceptable amplitude ranges.  Results of these pulser tests are presented 
in a later section.


\section{Instrument performance}

The ARA testbed installation was completed in the third week of
January 2011, at a location approximately 1.8~km grid East of the
IceCube detector, and has been operating nearly continuously since that
time. The instrument's internal Rubidium clock, stabilized
to a GPS receiver, provides a 1~Hz pulse that is used to trigger the
calibration pulser, and this in turns provides a stable repetitive
RF trigger which can be used to monitor trigger and amplitude
stability. The system also triggers on fluctuations of ambient
RF thermal noise at quiescent rates of typically 0.5-1~Hz, and in addition
we use the DAQ computer to force a trigger at 0.5~Hz to get
unbiased waveforms of the ambient RF noise in each antenna channel.
During the early part of the operation of the testbed in 
mid-January, we also took several dedicated runs where we turned on
the IceCube RF pulsers; since that time we have not repeated these
runs as it requires the presence of personnel who are not available
during the winter months at South Pole.
In the remainder of this section we report on various measures of 
the instrument performance for the first 4 months of 2011.

    \subsection{Environmental}

\begin{figure}[ht!]
 \begin{center}
\includegraphics[width=3.5in]{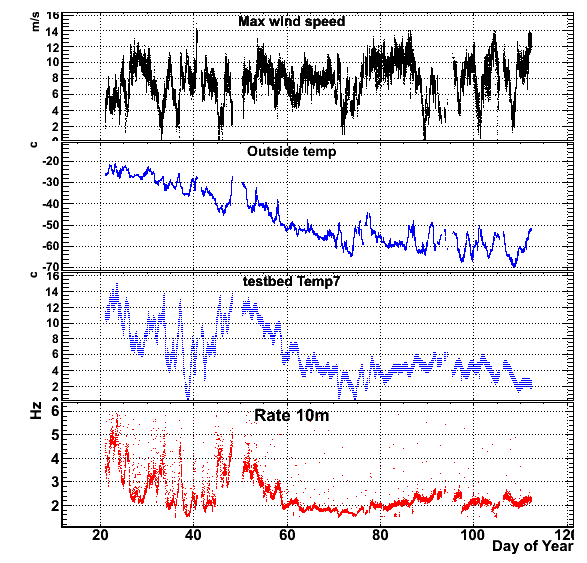}
 \caption{Top: Recorded South Pole wind speeds near the ARA-testbed site.
Upper middle: Air temperature near the site. Lower middle: internal temperature
with the data acquisition electronics housing. Bottom: Global trigger rate for
the ARA-testbed, showing a strong correlation with the electronics temperature. }
 \label{wind_temp}
 \end{center}
 \end{figure}

Fig.~\ref{wind_temp} shows a time series of several measured environmental
conditions for the ARA testbed. We monitored the wind speed and outside
temperature at locations near the South Pole station. Fig.~\ref{wind_temp}
shows that the overall average trigger rate of our system has remained quite low
for the duration of its operation so far, and shows no obvious correlation
with wind speed. Possible correlation of RF thermal noise with blowing snow was 
a hypothesis we wished to check, since blowing snow is known to 
create static charge buildup on capacitive metal structures in polar 
regions where the humidity is extremely low. For the ARA-testbed, all
surface equipment was deployed so that snow would drift in and bury it 
soon after it was deployed, to avoid having any remaining conductive structure on the
surface. We have so far found no evident trigger-rate correlation with wind speeds.

In the current electronics version, the digital-to-analog converter
(DAC) voltages that set the discriminator thresholds are held at
fixed digital values, but due to temperature-dependence, the actual output
voltages of the discriminator thresholds will drift slightly with temperature.
Our operational software has the capability to allow the threshold to servo
on the overall rates, but this was not implemented for the testbed yet, and 
thus we expect some correlation of the overall trigger rate with internal
temperature, and in fact we see such correlation clearly in the data.

    \subsection{Thermal noise floor}

\begin{figure*}[ht!]
 \begin{center}
\includegraphics[width=6.5in]{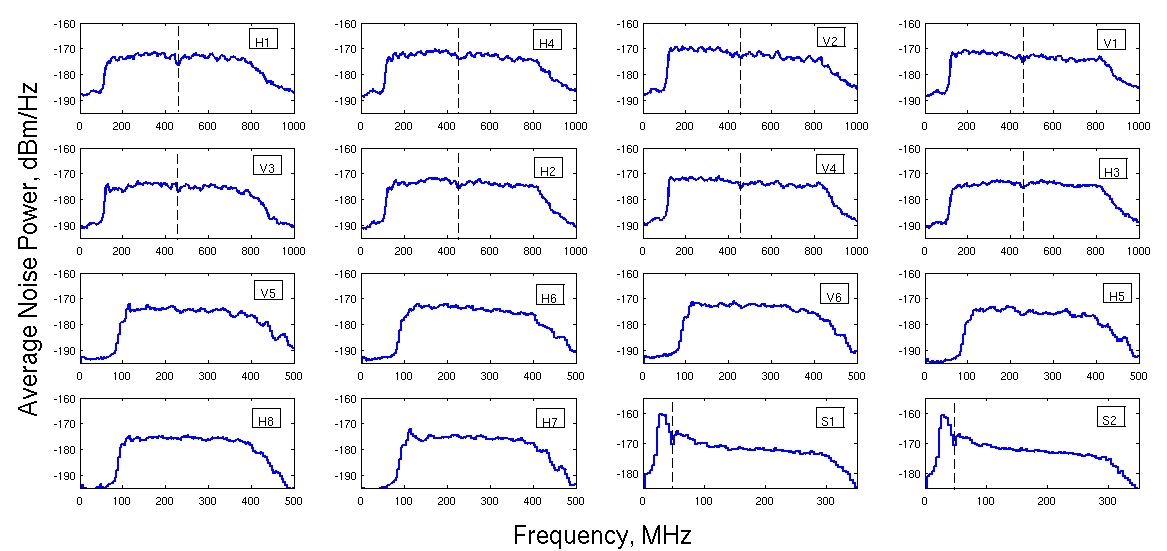}
 \caption{Average noise power spectral density for recent data for all antennas
in the current ARA-testbed.}
 \label{noisepower}
 \end{center}
 \end{figure*}

\begin{figure}[ht!]
 \begin{center}
\includegraphics[width=3.6in]{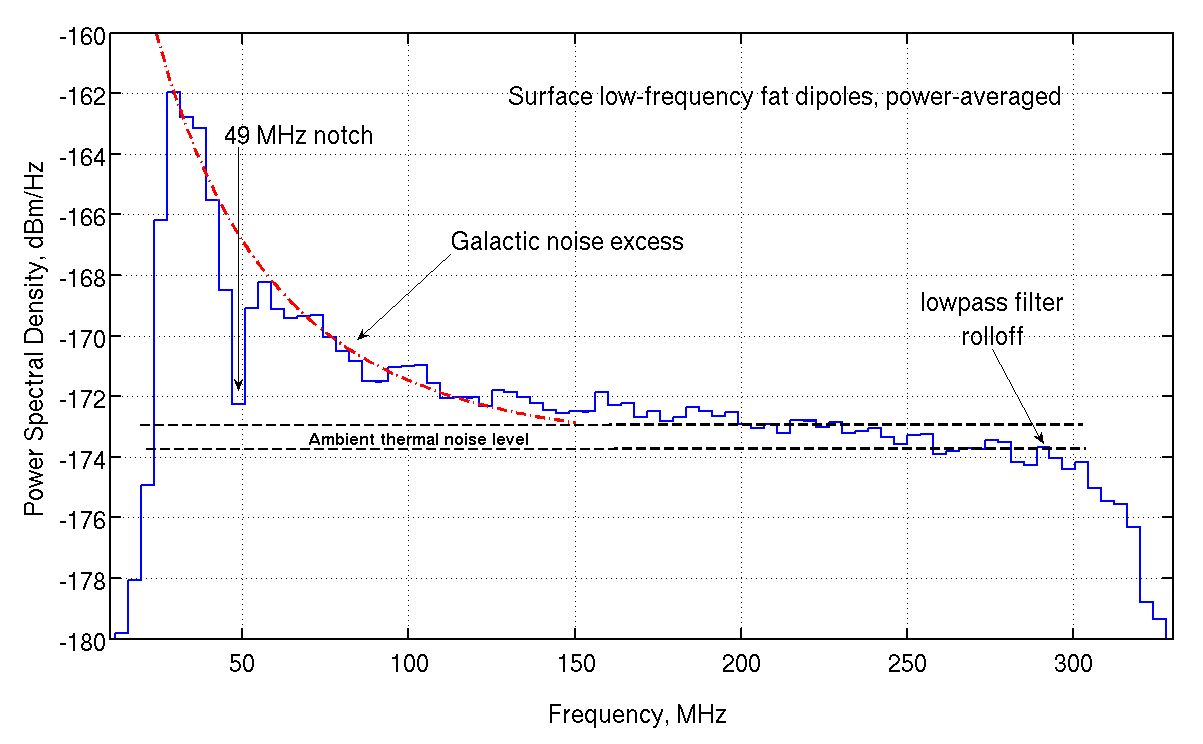}
 \caption{Average noise power spectral density for the surface antennas with
extended low frequency response.}
 \label{lowfreqnoise}
 \end{center}
 \end{figure}

An {\it in situ} estimate of the thermal noise floor for each of the 
antennas and receiver systems was made during the installation using
a bootstrap method. As noted above, the receiver noise figures were measured
separately, along with antenna transmissivity. During installation,
the antenna was replaced by a $50\Omega$ load termination at the ambient
temperature of the field site, and a series of noise sample waveforms were collected 
for the load as well as with the antenna replaced immediately afterwards. 
Because the termination gives only pure thermal Johnson noise
compared to the antenna, any change in the waveforms RMS noise once the
antenna is installed is due only to a change in the relative noise power,
and this gives the antenna temperature as compared to the load temperature.
In practice, the antenna temperatures observed in this calibration were
all comparable to the load temperature within the standard errors of the measurement,
which were of order 10\%. This shows that the antenna temperatures
are consistent with the ambient ice temperature. 
An exception in this procedure was observed for the two surface antennas,
which were subject to the strong Galactic noise component below 100~MHz. Above
100~MHz the observed noise in these antennas was consistent with the
thermal noise of the ice however.

Figure~\ref{noisepower} shows average Fourier power spectral density of 
the antenna+receiver thermal noise, taken for several hundred events
 which were selected as unbiased noise
sample waveforms from data taken in late April 2011, run 2533. The
thermal noise power level has been first-order calibrated, with a
systematic error of about $\pm 1$~dB. Pure thermal noise at 290~K produces
a noise power spectral density of -174~dBm/Hz, and in our case our average
ice+receiver thermal noise is just above this, at about 325~K, about
0.5~dB above room temperature equivalent. In most cases the data match
these expectations well, but in some cases our gain calibration appears to be
offset.

There are 
several features apparent in these data that are worth noting. 
First, for each channel the passband filter response is evident. For
the borehole antennas (upper eight), the passband is 130-850~MHz;
for the next 6 of 8, the passband is 100-400~MHz; these antennas are
near-surface antennas that are physically larger and thus respond
down to lower frequencies, although their beam patterns are still designed
to primarily view down into the ice. The last row contains
two borehole Hpol antennas, the QSCs, and the two low-frequency surface
fat dipole antennas. The QSCs do not turn on until about 200~MHz. 
Below their turn-on
frequency they act as effective terminations, so the thermal noise power
transitions over to Johnson noise. For the surface dipoles,
their LNAs are ineffective above about 300~MHz, so they are lowpassed
above that frequency.

\begin{figure}[ht!]
 \begin{center}
\includegraphics[width=3.5in]{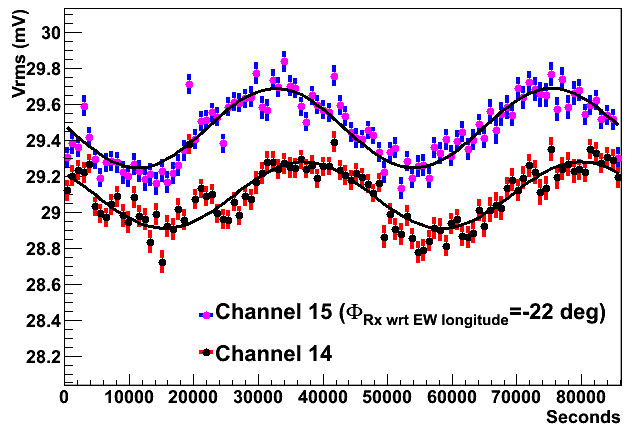}
 \caption{Modulation of the average RMS voltages observed in the
two surface low-frequency antennas during one
solar day. }
 \label{lowfreqmod}
 \end{center}
 \end{figure}

Second, some channels show ripple of up to $\pm 1$~dB ( $\sim \pm 10$\% in power) 
in the noise spectra; these are likely due to residual antenna impedance 
mismatches, which create low-level reflections between the antenna and LNA. 
Such effects can be mitigated with more careful matching which has been achieved
in other antennas that do not show these effects. In practice this spectral
ripple leads to very little distortion for pulse measurement since the 
reflection is causally late compared to the signal.

Third, in each
of the borehole antennas (the upper eight spectra), a dashed vertical line
marks the location of the notch filter, which is weakly visible in these
spectra since some of the antenna power is blocked at 450~MHz frequency.
In the lower antenna channels, the receivers had a 400~MHz lowpass filter
so the spectra are plotted only up to the point where they turn off.
In the two surface antennas (last two panes at lower left) there
is another notch at 46.28~MHz to suppress a known radar frequency;
this is also evident in the data. For these antennas,
the rapid rise of Galactic noise below 50~MHz is also evident.

\paragraph{Galactic Noise}
The average galactic radio noise begins to exceed that of the ambient
ice at roughly 150~MHz, rising quickly with decreasing frequency $\nu$ as
a power law with 
\begin{equation}
\label{galnoise}
T_{sky} = 800~{\rm K}~(\nu/100~{\rm MHz})^{-2.5}
\end{equation}
Fig~\ref{lowfreqnoise} shows the power spectral density for the
average of the two low-frequency antennas, showing the excess due
to Galactic noise at frequencies below 100~MHz. At 60~MHz, for example, 
the average sky temperature is about 3000~K, and we observe
a noise temperature about 6~dB above our thermal noise. For these antennas,
the LNAs had a receiver noise figure of about 100~K in this spectral range, and the
dipole response of the antenna effectively averages the noise power from
the sky above with that of the ice below with equal weight. Thus
we expect an antenna temperature of about 1700~K, a factor of about
5 above the antenna+system temperature due to the ice and LNA,
or 7~dB, consistent with expectations for galactic noise. 
The observed slope is also consistent with galactic noise;
at 100~MHz we see an increase of about a factor of 2 over
ambient ice+system temperature, consistent with equation~\ref{galnoise}
above.

Because the two surface low-frequency antenna are horizontal dipoles, we can
make use of the standard dipolar cosine response of the antenna beam pattern
to test for modulation of the galactic noise on the time scale of a sidereal 
day. As the dipole beam rotates through the galactic plane, which has an
inclination of $\sim 63^{\circ}$ with respect to the horizon at the South
Pole, the dipolar response pattern alternately views the hotter galactic plane 
and the cooler polar regions. We have tested for this effect for three months of
our testbed data, and these are folded together according to a sidereal period.
This is shown in Fig.~\ref{lowfreqmod} for the two surface antennas,
and there is clear sinusoidal modulation evident in the total power, which in this 
case has been low-pass filtered below 70~MHz to enhance the galactic noise component.
The nominal azimuthal orientation of these antennas differs by about
$22^{\circ}$ according to our as-built surveys, and we thus 
expect a phase shift between the two sinusoids. In fact we observe a $36^{\circ}$
phase difference, somewhat different than our expectations. However, given that
there are coaxial cables running within several meters of one of the antennas,
well-within the Fresnel zone of wavefront disturbance for these frequencies,
we attribute the difference to a distortion of the main beam of one of the
antennas. These results in any case lend confidence to our estimates
of the system and antenna temperatures for the ARA-37 testbed.

\subsubsection{Solar Radio burst of Feb. 13, 2011}

\begin{figure}[ht!]
 \begin{center}
\includegraphics[width=3.5in]{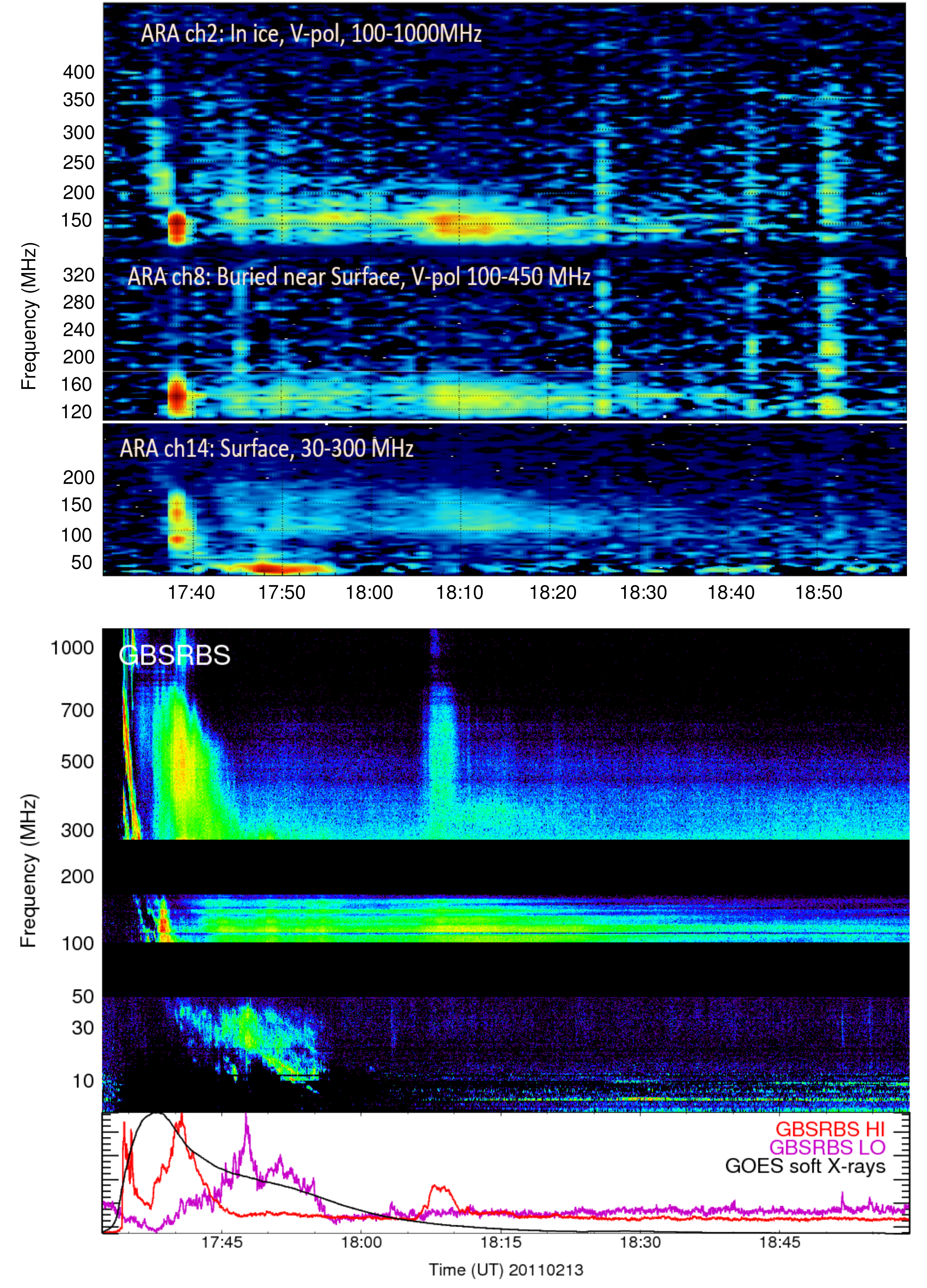}
 \caption{Solar radio burst of Feb. 13, 2011 as observed by ARA and
Green Bank Solar Rado Burst Spectrometer (GBSRBS). Top: dynamic spectra of 
relative RF power in several of the ARA antennas as a function of UT for that day
(arbitrary units); Bottom panels: dynamic spectra and time series plots of the
radio intensity seen at Green Bank.}
 \label{radioburst}
 \end{center}
 \end{figure}

On February 13,  2011, the Green Bank Solar Radio Burst Spectrometer
detected a series of strong type II solar radio bursts in a broadband measurement
spanning 10~MHz up to 1~GHz
~\cite{GBSRBS}, as seen Fig.~\ref{radioburst} (Bottom panels). 
Fig.~\ref{radioburst}(Top)
shows the resulting ARA dynamic spectra for several of our antennas for the
period of the burst. It is evident that the ARA testbed saw several
components of the burst sequence clearly in several different frequency bands. 
The observed spectra in ARA must be deconvolved from the frequency-dependent
response of the antennas to produce dynamic spectra with an absolute flux
density calibration, and this is outside the scope of the current report,
but  the results demonstrate the sensitivity of the ARA system, and also the
low background noise environment of the South Pole ARA site. Further work
is underway to produce interferometric images of the Sun during this event.
Fortunately, such events are relatively rare and will not pose a significant
noise background to ARA observations.

    \subsection{Radio-frequency Interference}

As we have noted above, understanding the radio-frequency interference environment
of the South Pole locale is critical to the effective design and operation
of a radio array in its vicinity. The ARA-testbed was deployed in the latter
part of the austral summer, and operations at Amundsen-Scott station continued
at full pace for some weeks after we began taking data, giving us a full
exposure to the types of interference that are common during the austral
summer season there, when most major science activities are conducted.

\begin{figure}[ht!]
 \begin{center}
\includegraphics[width=3.6in]{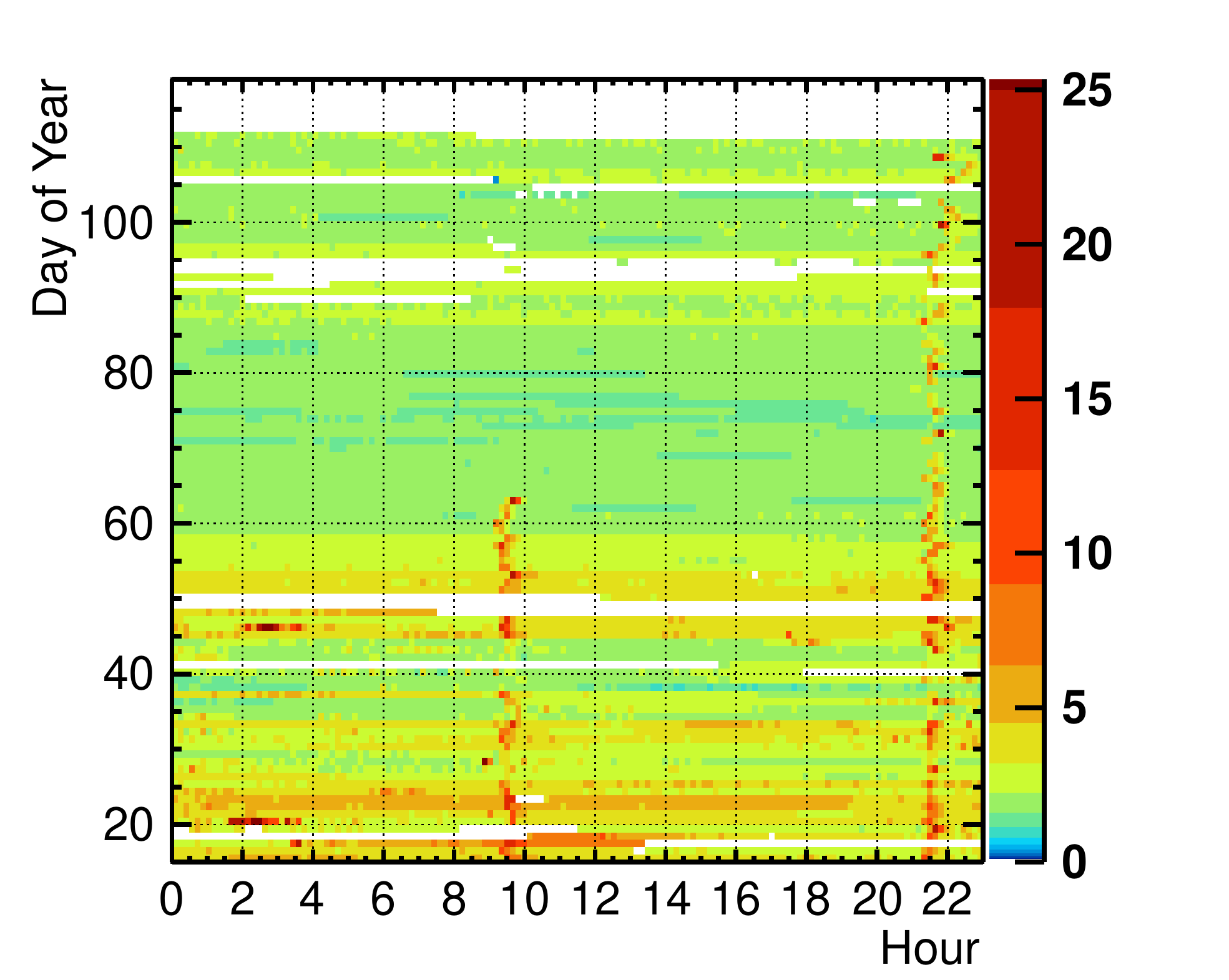}
\includegraphics[width=2.9in]{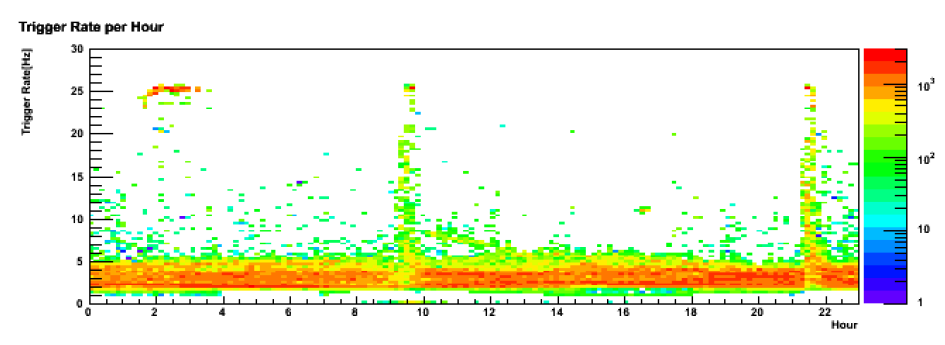}
 \caption{
Top: Hourly trigger rates for ARA prototype for about 3 months of operation of 2011. White space
in the plot indicates regions where stored South Pole data could not be transferred yet due to satellite
telemetry bandwidth restrictions.
Bottom: the daily profile of triggers, showing the peaks associated with weather balloon flights.
}
 \label{ARAsummary2}
 \end{center}
 \end{figure}

Figure~\ref{ARAsummary2} shows results of the trigger rates as a function of time for
the ARA-testbed over the entire quarter of operation. Not surprisingly, the austral summer
shows higher levels of sporadic interference. There are several specific periods where
interference is strong enough to significantly impact our operation: during 
meteorological balloon launches, which are conducted twice daily until early March,
when we observe the frequency reduced to once daily. These balloon launches
utilize a $\sim 400$~MHz transponder for data telemetry, and this produces strong
interference for us for about 1/2 hour in each flight.  During that time, our system
event rate is saturated, and the deadtime is nearly 100\%, but
this leads to a total loss of livetime of order 5\% , and half of this in the 
later season where the launches are reduced to one per day.

The other significant cause of strong interference is incoming or departing
aircraft which use a 129.3~MHz communication channel. In our current system,
while our filters are rolling off at that frequency, they are not strong enough
to suppress this signal below the level where it causes a high trigger rate, and
again we experience 100\% deadtime for a period of typically 20-30~minutes around the time
of arrival or departure of the flights. During peak activity periods, several
flights a day may land or depart, leading to an additional 5\% per day loss
of livetime on average. 

Other than these two sources, there are occasional sporadic interference episodes
associated with other activity, but these have had only a minimal effect on the
livetime to date. 

Figure~\ref{ARAsummary2} (top) also shows how the trigger rate has slowly declined
since the summer peak; this is likely not due to decreasing interference, but rather
the temperature dependence of the trigger thresholds, as discussed above.
Overall the RFI environment in the locale near Amundsen-Scott station is
quite acceptable for our requirements, and we estimate the yearly average 
loss of livetime due to interference at of order 3\% at most.

In addition to assessing noise generated by equipment already in place at the station,
we installed a wind turbine system at a distance of order 800~m from the
ARA-testbed as part of a program to investigate 
autonomous long-term power sources (detailed in Appendix A) for
ARA-37 and future arrays that are deployed further out from the existing
South Pole infrastructure. We have searched our data for triggers which
specifically reconstruct toward the direction of the wind turbine, and we
find none, indicating that this technology will likely be acceptable for use
for ARA.

    \subsection{Calibration Pulser}

\begin{figure}[ht!]
 \begin{center}
\includegraphics[width=3.6in]{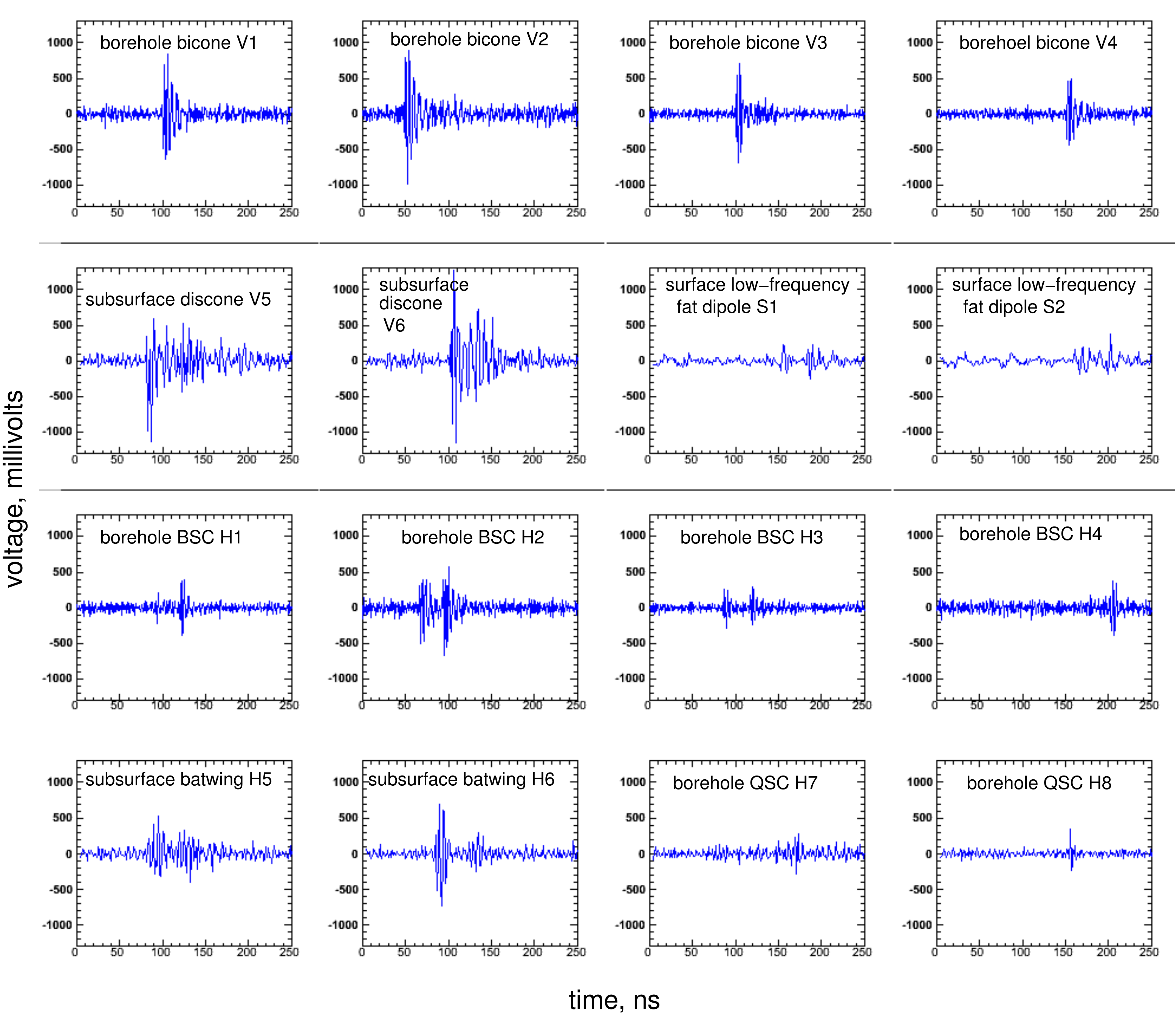}
 \caption{Local calibration pulser event waveforms in all ARA-testbed antennas.}
 \label{Calpulser1}
 \end{center}
 \end{figure}

The local calibration pulser, which pulses at a repetition rate of 1~Hz
during all periods when the ARA-testbed system is running, provides a
continuous measure of the trigger functionality, and a stable timing 
and amplitude reference for monitoring of the ongoing performance of 
the system. An example of an individual calibration pulser event is shown in
Fig.~\ref{Calpulser1}. Because the calibration pulser is located at nearly
the same depth as most of the borehole antennas, the signal shape 
as seen by those antennas has the broadest bandwidth and thus tends to
be very narrow and coherent in time. Because the radius of the calibration
pulser borehole is only three times the $\sim 10$~m 
radius of the inner borehole array, an amplitude gradient across the array
is quite evident in the data. 

We also noted in these data that there is a second pulse, which we 
hypothesize is a reflection, appearing in the Hpol antennas only. Other
tests done during deployment using different calibration locations
saw no similar reflections, so it does not appear to be something intrinsic
to these channels.
The region where we deployed the ARA-testbed is still relatively close to the
original station established in the 1950's. The depth at which we deployed
our antennas, typically 20-30~m, is in fact the depth at which the
1950 surface layer is found. Because many air-drops of equipment in the
vicinity of the Pole were done during this period, and much equipment was
left in the field, it is possible we have detected a horizontal metallic
object within several meters of our boreholes, and we are investigating this
further in the data.

\begin{figure}[ht!]
 \begin{center}
\includegraphics[width=3.35in]{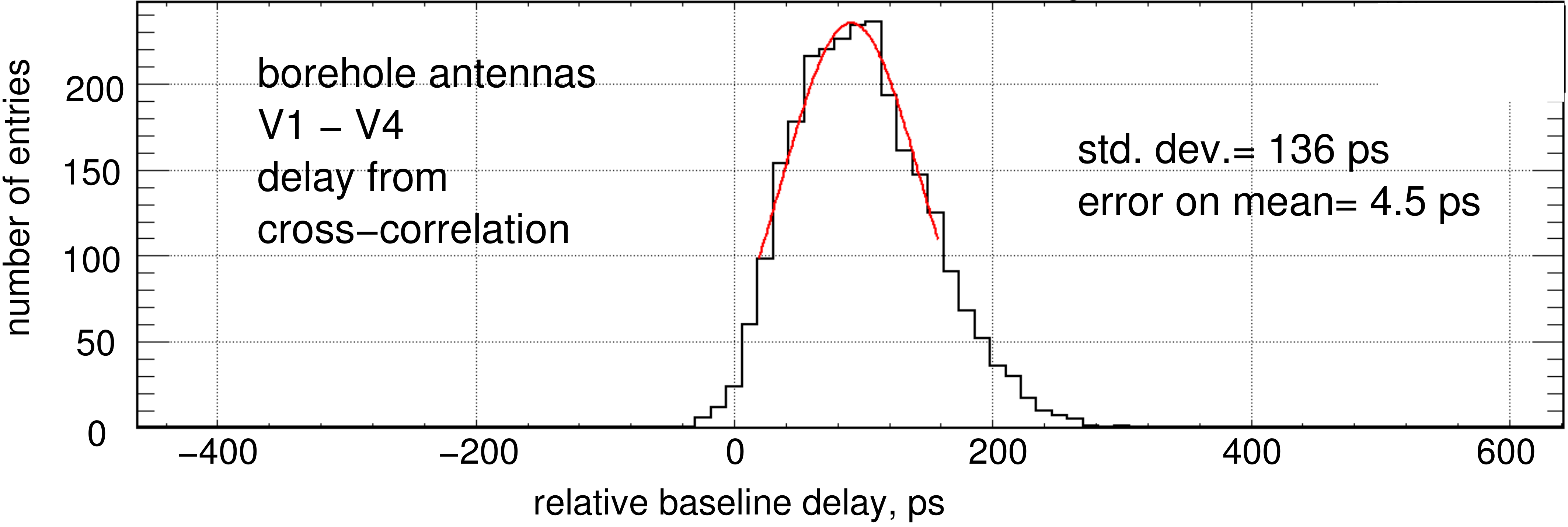}
 \caption{Example of delay histogram from cross-correlated waveforms
for an ensemble of about 2600 calibration
pulser events, in this case for borehole
antennas V1 and V4. The origin is arbitrary in this case.}
 \label{CP2delay}
 \end{center}
 \end{figure}

One of the important functions of the calibration pulsers is to
provide a stable reference signal that can be used to determine the
geometry of the array {\it in situ}, via the relative delays between
the received waveforms within the array. Fig.~\ref{CP2delay} shows
a typical example of a histogram of the delays between Vpol
borehole antennas V1 and V4 for an ensemble of about 2600
calibration pulser events. The delays for each event are extracted by
cross-correlating the received waveforms for each event, and this
delay distribution is shown in the plot, with a fitted Gaussian to
the peak region giving a well-defined mean value (arbitrary in this case)
with an estimated error of about 4.5~ps, and a Gaussian standard deviation of 136 ps. 
The standard error on the fitted mean represents the ultimate limit of the
calibration pulser data for determining geometry within the array --
in this case the 4.5~ps resolution corresponds to about 1~mm in ice.
Thus the timing resolution is more than adequate for calibration; our
reconstruction requirements for the geometry can easily tolerate few cm
accuracy on locations.

The standard deviation of the fitted Gaussian for the pulser
delay distribution also provides a calibration of the potential
angular resolution for the system. A standard deviation of 
136~ps in this case corresponds to about 2.7~cm in the ARA-testbed ice.
Our angular reconstruction of signal arrival directions depends on
the ratio of our standard error on the waveform arrival to the
length of the projected baseline between two antennas, eg:
$$\Delta \theta_{ij} = \frac{c}{n}\frac{\tau_{ij}}{\vec{S}\cdot \vec{B}_{ij}} $$
where $\tau_{ij} = t_i - t_j$ is the baseline delay for wavefront arrival
times $t_i$ and $t_j$ at antennas $i,j$, and 
$\vec{S}$ is the Poynting vector of the incoming wave, here 
assumed to be a plane wave for simplicity. For typical
baseline lengths of 15~m or more, the implied angular resolution
per baseline is of order $0.1^{\circ}$ for 136~ps timing error,
again more than adequate for our requirements.

\begin{figure}[ht!]
 \begin{center}
\includegraphics[width=3.35in]{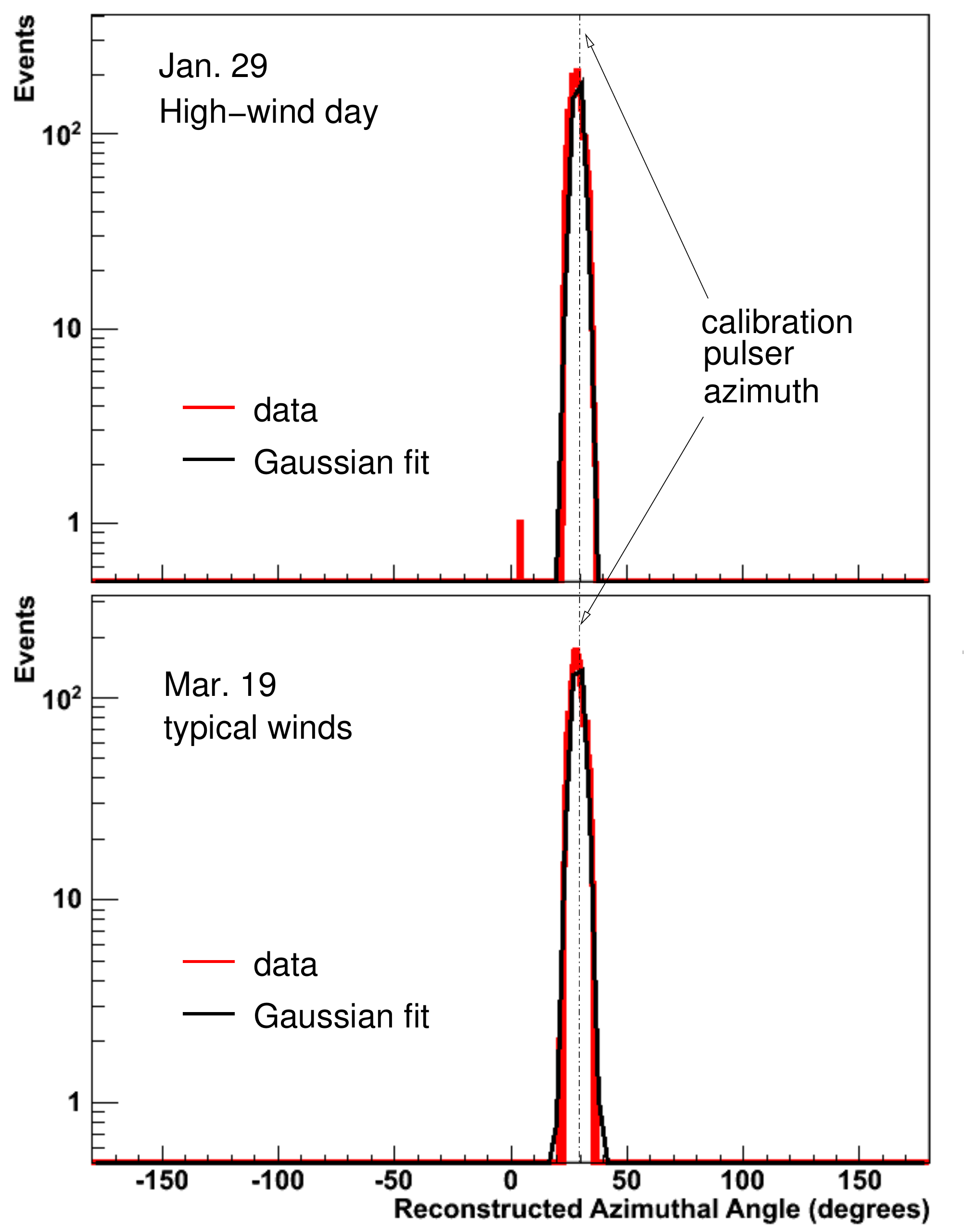}
 \caption{Examples of the azimuth distribution of events observed
during two days of the ARA-testbed operation.}
 \label{jan29_mar19}
 \end{center}
 \end{figure}

\begin{figure}[ht!]
 \begin{center}
\includegraphics[width=3.35in]{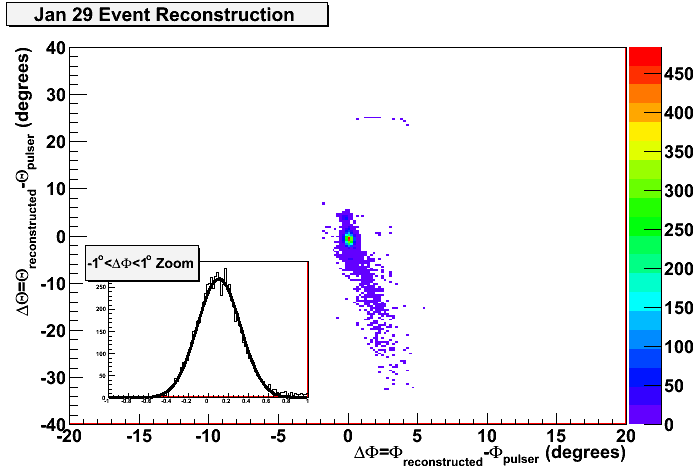}
 \caption{Two-dimensional elevation vs. azimuth reconstructed distribution of events 
referenced to the calibration pulser direction, and an inset that shows the
azimuth distribution, with angular resolution well under 1\deg.}
 \label{jan29}
 \end{center}
 \end{figure}

Fig.~\ref{jan29_mar19} shows the distribution of the
reconstructed azimuthal direction of events observed on two separate typical
days during the operation period. Events that RF-triggered the testbed during
these periods were either thermal noise fluctuations, which can 
form random coincidences but do not reconstruct as plane- or spherical-waves
when analyzed offline, or RF triggers from the calibration pulser, or from
any other RF interference or backgrounds that may be present. The latter
two types of events recur from fixed directions: the calibration pulser from
a fixed $\sim 30^{\circ}$ azimuth; RFI can occur from a variety of directions
but will be associated with the South Pole Station, and are likely to repeat,
thus giving clusters of events in stable directions associated with human activity.
It has also been hypothesized that wind-blown snow could create electrical discharges
itself, or by depositing charge on structures which then discharge once they reach
a high-enough potential. Thus we have looked for a background of random triggers
that are associate with higher-than-normal winds, or structures such as our wind-turbine
system.

We find that almost all RF trigger events observed have
an azimuth consistent with that of the calibration pulser with respect to the
center of the array.  January 29, 2011 was a period of 
relatively high winds ($>20$~kt), which produces more opportunity for
static discharge on any above-surface structures, 
and also yields higher power generation in the nearby wind turbine system.
By contrast March 19 was a period of relatively calm wind. In neither case
did we observe any significant number of events that 
reconstructed to any other direction, including the $\sim 155^{\circ}$ azimuth
of the nearby wind turbine system, and other directions associated
with the IceCube Laboratory and South Pole Station. Fig.~\ref{jan29}
shows a two-dimensional map of the reconstructed events for the
higher wind date, with an inset showing the azimuth distribution.
A very small fraction of events misreconstruct in elevation angle
as evident in the plot, but the vast majority of events are
consistent with the calibration pulser only.

    \subsection{Deep pulser measurements}

In our measurements, all of the IceCube pulsers show some degree of saturation
in the received signals at the ARA-testbed. While this bodes well for future
studies of long-range ice transmission for ARA, the saturated waveforms in the
ARA-testbed complicate efforts at precision estimation 
ice transmission parameters.

\begin{figure}[ht!]
 \begin{center}
\includegraphics[width=3.6in]{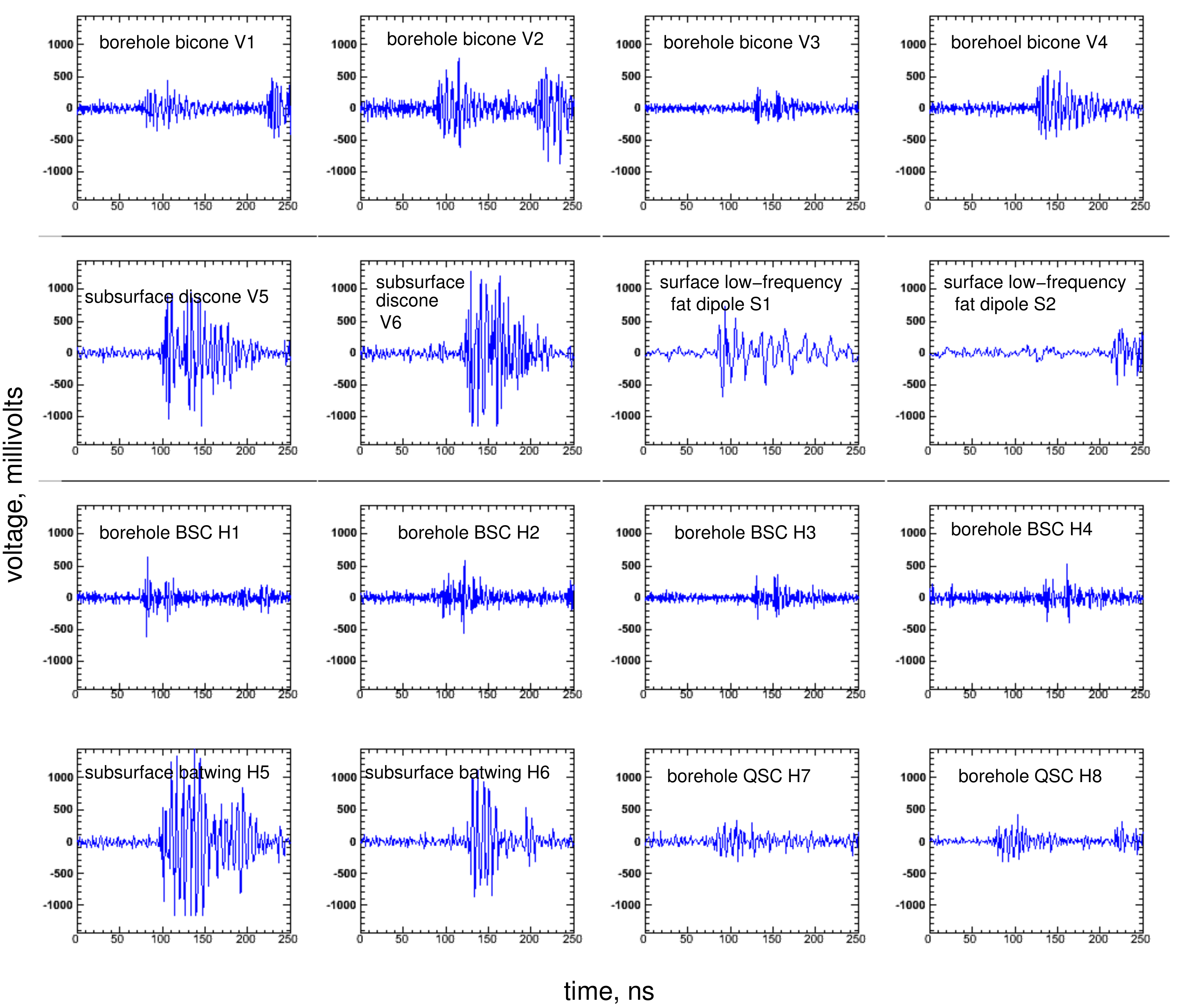}
 \caption{Deep calibration pulser event waveforms in all ARA-testbed antennas.}
 \label{deeppulser1}
 \end{center}
 \end{figure}

In addition, the bicone antenna design for deployment concentric to the IceCube
cables, and under the extreme high pressure of the re-freeze that occurs in the
hot-water-filled IceCube boreholes, led to some challenges in the 
antenna construction that also came with risk. For example, the clam-shell
jointing of the antennas to surround the IceCube cable must be done during 
a rapid deployment, and inadequate contact for the joint can lead to 
shunt-capacitance across the slot, and cause some radiation of Hpol as well
as Vpol signals. Also, the long-wire vertical IceCube cable in close proximity to
the antenna can lead to re-radiation and scattering of the signal. 
As a result of analysis of the signals and comparison with some prior
lab measurements, we conclude that the transmitted pulse from the deep
pulser was dispersed in time over typically a $\sim 25$~ns time window,
although it appears that most of the pulse power was in fact contained withing
this dispersed pulse.

Despite these difficulties, we have observed signals from these transmitters
at up to a 3.2~km slant range from the source. The best example of this is shown
in Fig.~\ref{deeppulser1} which comes from the furthest and deepest of the
deployed pulsers, installed in IceCube Hole \#1 at a depth of 2450m 
and a horizontal distance of about 2~km from the ARA-testbed,
for a total slant range of 3.16~km at an angle of about 40$^{\circ}$ from the 
vertical. Some of the observed signals still saturate our DAQ dynamic range
even at this distance,  and we do observe significant
Hpol signal as well as Vpol; suggesting that the clam-shell coupling of the
antenna was not complete. The Hpol signals also have higher frequency content
than the Vpol signals on average, and are factors of 2-3 weaker in
amplitude, which is consistent with radiation from partial slots 
along the antenna's vertical joint. In several cases, receiving antennas are partially 
shadowed by antennas in the foreground with respect to the pulse arrival
direction. The radius of the first Fresnel zone is the region around
any obstruction within which a propagating wave is disturbed, and for 
the ARA-testbed antennas for a source
at distance $R_s$ and the foreground antenna at distance $d$ from the receive
antenna we have
\begin{equation}
R_F = \sqrt{\frac{\lambda~ R_s~ d}{R_s + d}}.
\end{equation}
For $R_s = 3200$~m, and $d \sim 20$~m, $R_F \sim 4$~m for $\lambda \sim 1$m in ice,
indicating that we would expect several of the antennas to suffer from
significant shadowing, which we observe.

\begin{figure}[ht!]
 \begin{center}
\includegraphics[width=3.5in]{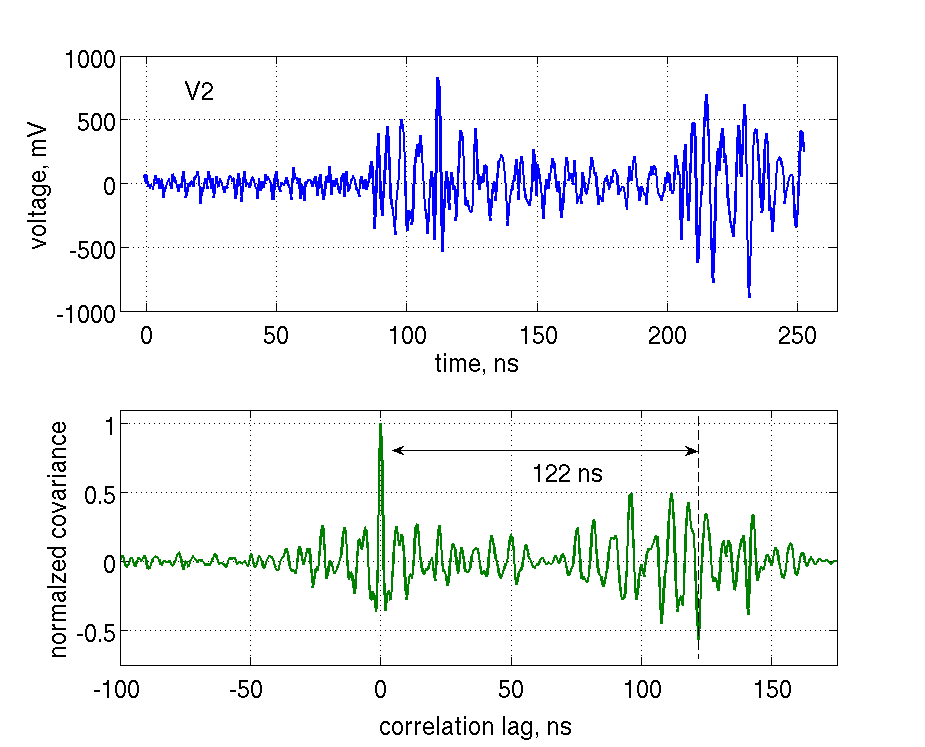}
 \caption{Top: Antenna individual V2 received waveform from the deep pulser,
showing the secondary surface reflection near the end of the trace. 
Bottom:
cross-covariance of the first pulse with the entire waveform. 
Note that the peak covariance from the second (reflected) pulse is inverted
in phase with the primary pulse, at a lag time of 122~ns, which is within 1\%
of the propagation time estimate for the reflected ray.}
 \label{surf_refl_zoom}
 \end{center}
 \end{figure}

\subsubsection{Attenuation length estimates}

If we isolate our attention to the foreground Vpol antennas that are undisturbed,
we may use the resulting amplitude to estimate the average attenuation length
$\langle L_{\alpha} \rangle$ 
over the entire 3.16 km propagation path, along
with our knowledge of the antenna and pulse parameters at the source. 
The attenuation length can be determined from the Friis equation~\cite{Kraus},
which relates the received power $P_{r}$ to the transmit power $P_{t}$ as
a function of frequency:
\begin{equation}
P_{r}(\nu) = 4 \pi~  P_{t}(\nu)~ e^{-2R/\langle L_{\alpha}\rangle}~~ \frac{\nu^4 ~A_{r}~ A_{t}}{c^2~ R^2}
\end{equation}
where $R$ is the path distance through the ice, and $A_{r}, A_{t}$ are the
receive and transmit antenna effective areas (including antenna coupling efficiency), 
which are in general also functions of
frequency $\nu$. Inverting this for the average attenuation length we have:
\begin{equation}
\langle L_{\alpha} \rangle = -2R \left ( \log \left [ \frac{P_{r}}{P_{t}} 
\left (  \frac{c^4~ R^2}{4\pi~ \nu^4 ~A_{r}~ A_{t}} \right ) \right ] \right )^{-1}~.
\end{equation}
The quantity $\langle L_{\alpha} \rangle$ in this case is a weighted average,
where the weights are determined by the vertical temperature profile of the
ray paths, since the attenuation coefficient (in units of m$^{-1}$)
is $\alpha = L_{\alpha}^{-1}  = \alpha(\nu,T_{ice})$, a strong function of
the ice temperature $T_{ice}$. 

\begin{figure}[ht!]
 \begin{center}
\includegraphics[width=3.5in]{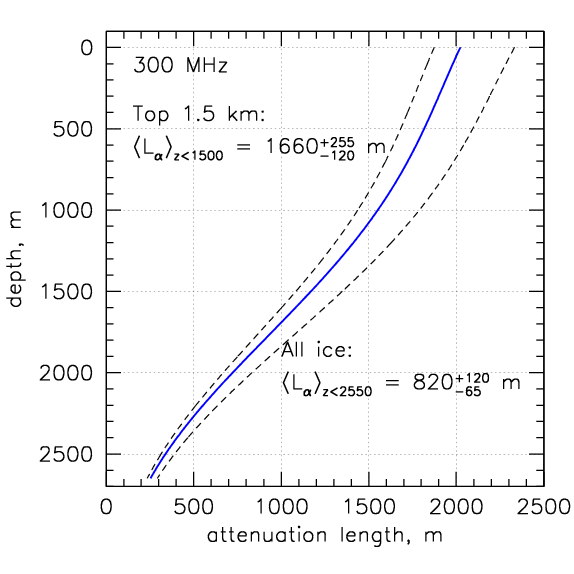}
 \caption{The field attenuation length over the 3.16~km path from the deep
pulser to the ARA testbed as a function of depth. The average attenuation
length over all depths is also given.}
 \label{L_v_depth}
 \end{center}
 \end{figure}

Accounting for the transmitted power, for the bicone transmitter
antenna response, estimated insertion losses, antenna beam pattern,
a first-order estimate of the signal saturation,
and the receive antenna response, our estiamte of the attenuation length over the
referenced to 300~MHz is
plotted in Fig.~\ref{L_v_depth} as a function of depth. The depth dependence
is deconvolved from the total attenuation under the assumption that the
depth dependence is dominated by the ice temperature profile, which has
been determined quite well for South Polar ice. The average attenuation
length for all depths, $820 (+120,-65)$~m is also given.
The errors here are dominated by our systematics on the received amplitude and
pulse shape. The resulting value is quite consistent with
values derived from vertical bottom-bounce methods at the South Pole~\cite{icepaper},
and the systematics here are different from those in the bottom-bounce
measurement in which the bottom reflection coefficient is unknown. 
Note that the attenuation length derived is dominated by 
the extended propagation distance in the relatively
warm basal ice; considering the upper 1.5~km where the ice is much
colder than the basal ice, the effective average attenuation length is nearly 1.7~km.
This remarkably long field attenuation length of the ice in this extremely cold
region of the cryosphere sets it apart from any other natural (and in fact most man-made)
dielectrics as the most transparent solid medium on Earth. It is
perhaps the single most important characteristic of the ice for
enhancing the performance and viability of an array such as ARA.

\subsubsection{Surface reflection}

\begin{figure}[ht!]
 \begin{center}
\includegraphics[width=3.5in]{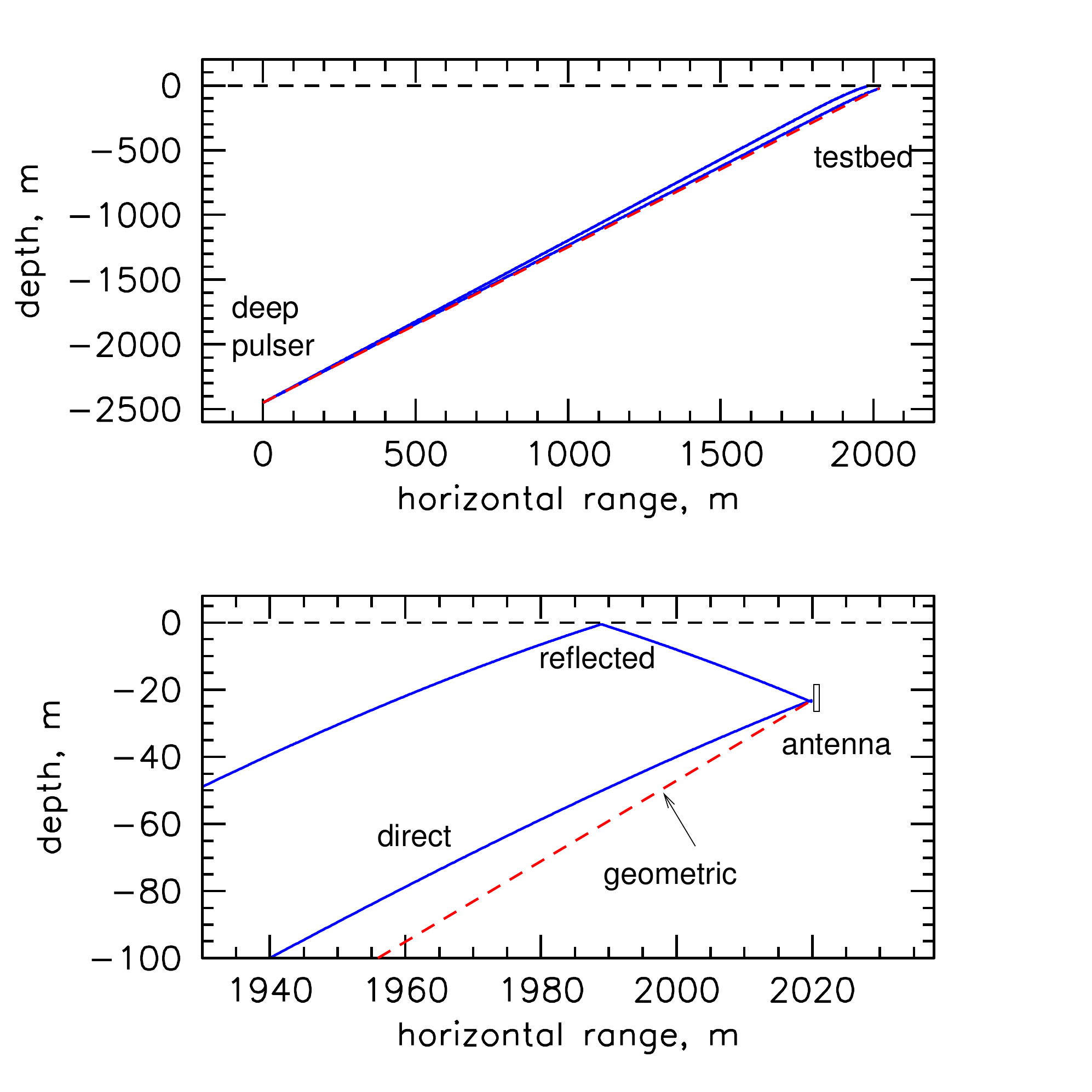}
 \caption{Ray paths for direct and reflected rays to antenna V2. The red-dash
line gives the geometric ray path for the direct ray.}
 \label{surf_refl}
 \end{center}
 \end{figure}

Another interesting feature observed in the deep pulser signals,
which arrive at the ARA-testbed as a close approximation to a plane wave,
is the surface reflection which is seen in several antennas, most clearly in
borehole bicones V1 and  V2. These differ in depths by several meters,
leading to a different delay time before the reflection arrives. In 
antenna V2, the reflection is clearly phase-inverted compared to the primary
signal; V1's record cuts off too early to observe this clearly.
To verify the presence of the reflection, the waveform (top) and a cross-covariance
estimate (bottom) of the primary and secondary pulses with the primary pulse are shown
in Fig.~\ref{surf_refl_zoom}. The cross-covariance has its highest correlation
with the primary pulse at a 122~ns delay, where the peak is a 60\% anticorrelation,
as seen in the Figure.
V2's signal amplitude for the reflection is also quite close to
the direct signal amplitude, which is consistent with a total
internal reflection off the top of the ice, as expected for the estimated
$57^{\circ}$ incident angle of arrival with respect to the ice surface
normal.

Fig.~\ref{surf_refl} shows numerical ray trace paths for this antenna from
the IceCube deep pulser (top), and a zoom to the region of the reflection
in the bottom pane. The numerical ray tracer uses the eikonal equation and
a refractive index model developed based on radio index of refraction data
from the RICE experiment at the South Pole, and returns both the ray
path and total wave transit time along the ray. The delay observed in
antenna V2's waveform from the direct to inverted reflection is 120.0~ns.
For the ray paths from Fig.~\ref{surf_refl}, the direct path transit
time is 18,724.53 ns, and the reflected ray path requires 18,669.92 ns to the
surface, followed by 176.6 ns from the surface to the antenna, for a 
relative delay of 122~ns with respect to the direct path, a result 
identical to that observed in the data. This also independently confirms the
precision of our index-of-refraction model.

It is a generic feature of a subsurface radio array that signals
may be observed both directly and via the surface reflection, or even
via near-reflections for rays that are curved back down near the surface; this effect
is not yet incorporated into our simulations, but will improve the
detection efficiency in general, and give some additional reconstruction
power as well.
 

\section{Expected Performance for ARA-37}

We characterize the expected performance of ARA based on an extensive suite of
Monte Carlo and other simulation and modeling tools developed over the
last two decades. Several of the investigators on this proposal have
been involved in radio methods for detection of UHE neutrinos since the
mid-1990's and thus the heritage of our simulation methodology is many
generations deep and has been proven over a wide range of active and completed
experiments, including the RICE~\cite{RICE06} and AURA~\cite{AURA} 
experiments, which are 
direct pre-cursors to ARA, the Goldstone Lunar Ultra-high energy neutrino 
Experiment (GLUE)~\cite{GLUE04}, and the ANITA experiment~\cite{ANITA2}.

\begin{figure}[tb!]
  \includegraphics[width=3.2in]{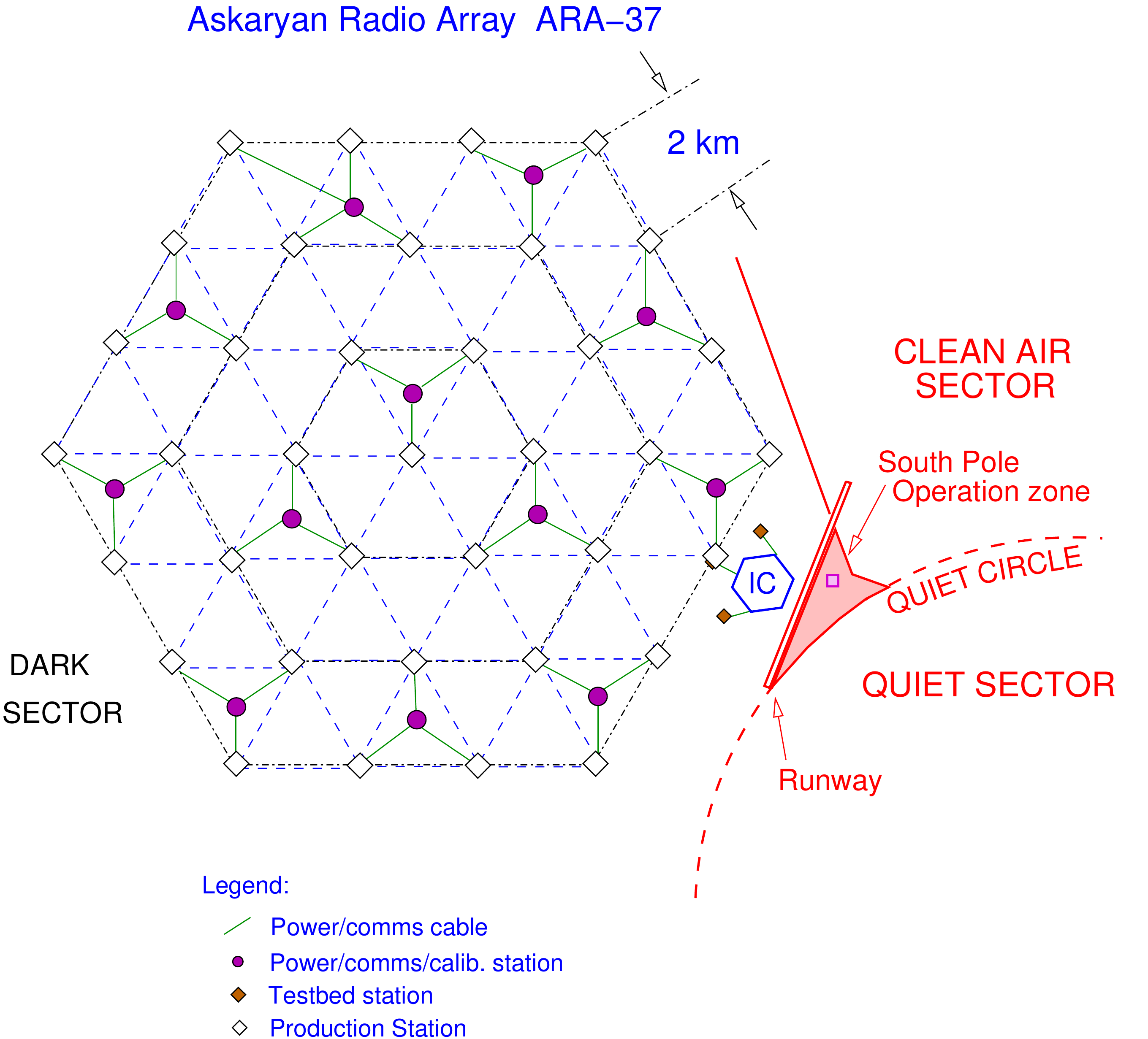}
\begin{small}
\caption{ Planned layout of the 37 ARA stations with respect to the
South Pole Station and associated sectors.
    \label{geom}
  }
\end{small}
\end{figure}

Our Monte Carlo tools include detailed ice attenuation modeling and raytracing
to account for the gradient in the index of refraction of the ice vs. depth. They
provide state-of-the-art modeling of the Askaryan radiation from showers via
tested parametrizations, which have been validated by direct measurements of
the Askaryan effect in ice at SLAC~\cite{slac07}. Neutrino propagation
through the earth and ice sheets is modeled in detail, and the particle
physics of the interaction, including neutral and charged-current effects,
fully-mixed neutrino flavors, and secondary shower production due to
charged-current $\tau$- and $\mu-$leptons are accounted for in the models.
Finally, the detectors are also modeled with high fidelity, including 
the effects of Rician noise in the detection process, spectral response
functions of the antennas, and full 3-D polarization propagation for the
radio waves that interact with the detector. We thus have reason to report
these performance estimates with some confidence.

\begin{figure}[tb!]
  \includegraphics[width=3.4in]{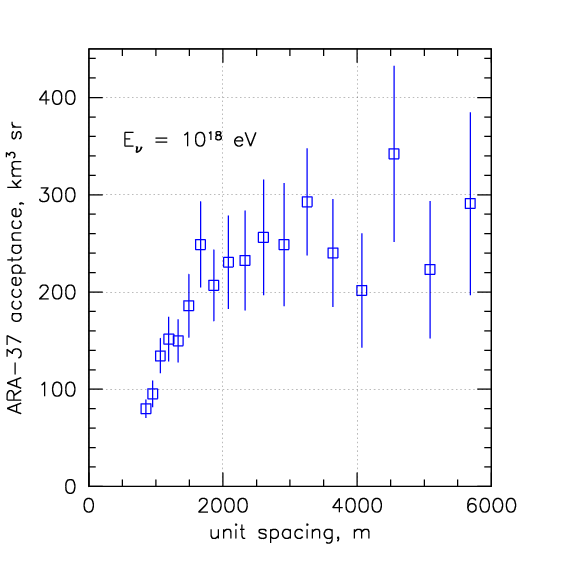}
\caption{  Simulation of the effective acceptance for ARA-37 as a
function of the inter-station spacing, for an energy of 1~EeV.
    \label{ara_spacing}
  }

\end{figure}

  \subsection{Station Spacing}

In initial studies of the ARA-37 array~\cite{IceRay09}, we adopted a 1.33~km
spacing between stations as a compromise which gave adequate sensitivity
while still allowing for enough overlap between the stations' effective
target volumes to yield reasonable fraction of multi-station 
coincident events. However, in the interim, indications of possible heavier
nuclear composition for the UHECRs have led to estimates for the cosmogenic neutrino flux
which are significantly lower than for a pure proton spectrum, and thus
in our current designs, we optimize the ARA-37 array for discovery potential,
to maximize the number of detected events.  Fig.~\ref{ara_spacing} gives
the results of a parametric study of the neutrino acceptance of the array
as a function of the spacing for a neutrino energy of 1~EeV, which is
an excellent proxy on average for the total number of detected events
integrated over a typical cosmogenic neutrino spectrum. 

We find that the acceptance grows rapidly with spacing starting with
sub-kilometric sizes, and then
becomes fully saturated at about 2-3~km. In our current design we have
thus adopted 2~km as our baseline spacing; 
this choice reflects a good compromise between maximizing the
sensitivity while still recognizing the potential logistical costs that grow
with array size.

		\subsection{Sensitivity}

The primary metric for detection of cosmogenic neutrinos, which are presumed to
arrive isotropically on the sky, is the volumetric acceptance $\mathcal{V}\Omega$, 
in units of km$^3$ steradians. An equivalent parametrization is the
areal acceptance $\mathcal{A}\Omega$ (km$^2$ sr) and the two are closely related
by $\mathcal{A}\Omega = \mathcal{V}\Omega/ \mathcal{L}_{int}(E_{\nu})$ where
$\mathcal{L}_{int}$ is the interaction length of the neutrinos as a function of 
neutrino energy $E_{\nu}$. The volumetric acceptance, divided by the 
instrumented target fiducial volume, gives a measure of the detection efficiency of
neutrinos which interact within the fiducial volume of a detector.
In the case of ARA-37, one realization of the simulation uses a cylindrical ice target
volume of radius 10~km, and depth 2~km. Because of earth attenuation, neutrinos arrive
almost exclusively from above the horizon, nominally 
giving $\sim 2\pi$ steradians for the solid angle.
The net target acceptance of the simulated ARA detector is thus just over
4000 km$^3$ sr, and this represents the maximum neutrino volumetric acceptance the
simulation could obtain. However, once the constraints given by the neutrino interaction
cross section, the ice attenuation length, and the ray-tracing geometry of the
ice, and the antenna response functions are imposed, the effective acceptance
becomes an energy-dependent fraction of the initial target acceptance.

\begin{figure}[tb!]
\includegraphics[width=3.5in]{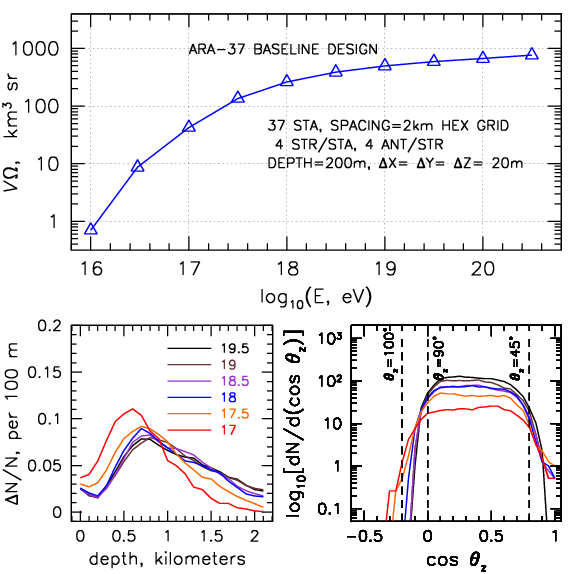}
\begin{small}
\caption{  Top (large pane): Simulated neutrino volumetric acceptance (km$^3$ sr water equivalent) for the ARA instrument baseline design. 
Bottom left: Depth
distribution of simulated events for different neutrino energies, showing the
contribution of deep ice down to 2~km or more at the higher energies.
Bottom  right: zenith angle distribution of detected neutrino arrival directions for a range of neutrino energies. Events are detected over a range from $\sim 45^{\circ}$
above the horizon to $\sim 5^{\circ}$ below it.
    \label{Veff11}
  }
\end{small}
\end{figure}

Fig.~\ref{Veff11} shows the simulated $\mathcal{V}\Omega$ results for our adopted baseline design, 
as a function of neutrino energy in the range
of interest for the cosmogenic neutrino flux. The acceptance reaches the level of
$> 200$~km$^3$~sr at the mid-range of the cosmogenic neutrino flux, which has a broad
plateau from about $5 \times 10^{17}$~eV up to just over $10^{18}$~eV, and continues
growing slowly up to the highest simulated energies, approaching a Teraton-steradian.

Fig.~\ref{Veff11}(bottom) gives a plot summary of some characteristics of the simulated data vs.
neutrino energy. On the bottom left, the depth distribution of detected events is shown
normalized to the event fraction per hundred meters. Events originating from below about
2~km depth tend to be suppressed, as the attenuation of the ice begins to grow quickly in the
warmer basal ice~\cite{icepaper}.
The estimated average attenuation length of our 2~km-deep fiducial 
volume is about 1.5~km, a factor of three better
than ice in locations such as the Ross Ice Shelf, where
the thickness is limited to several hundred m, and the attenuation lengths are
comparable to this thickness scale.
SP ice, especially in the upper 2 km
of its depth, is the clearest solid dielectric medium on Earth in the radio range,
and is the most compelling natural feature of the ARA site.

Fig.~\ref{Veff11}(bottom) also shows the
arrival zenith angular distribution of neutrino events that were detected,
showing that the neutrino angular acceptance spans a range from
$\sim 5^{\circ}$ below the horizon to $\sim 45^{\circ}$ above the horizon,
more than 6 steradians of solid angle. 


\begin{table}[hbt!]
\caption{Expected numbers of events $N_{\nu}$ from several UHE neutrino models,
comparing published values from the 2008 ANITA-II flight with predicted events
for a three-year exposure for ARA-37.
\label{nurates}}
\vspace{3mm}
 \begin{footnotesize}
  \begin{tabular}{lcr}
\hline \hline
{ {\bf Model \& references}}~~~~~~~~~{$N_{\nu}$}: &   ANITA-II,    &  ARA, \\ 
                                                   &(2008 flight)& 3 years \\           \hline
{\it Baseline cosmogenic models:} &  &  \\
~~~~~~Protheroe \& Johnson 1996~\cite{PJ96} & 0.6  &  59   \\
~~~~~~Engel, Seckel, Stanev 2001~\cite{Engel01} & 0.33  & 47   \\
~~~~~~Kotera,Allard, \& Olinto 2010~\cite{Kotera2010}& 0.5  & 59 \\
{\it Strong source evolution models:}&  &  \\
~~~~~~Engel, Seckel, Stanev 2001~\cite{Engel01} & 1.0 &  148 \\
~~~~~~Kalashev {\it et al.} 2002~\cite{Kal02} & 5.8 & 146 \\
~~~~~~Barger, Huber, \& Marfatia 2006~\cite{Barger06} & 3.5 & 154  \\
~~~~~~Yuksel \& Kistler 2007~\cite{Yuksel07} & 1.7 &  221  \\
{\it Mixed-Iron-Composition:}&  &  \\
~~~~~~Ave {\it et al.} 2005~\cite{Ave05}~~ & 0.01 &  6.6 \\ 
~~~~~~Stanev 2008~\cite{Stanev07} & 0.0002   &  1.5 \\
~~~~~~Kotera, Allard, \& Olinto 2010~\cite{Kotera2010} upper& 0.08  &  11.3 \\
~~~~~~Kotera, Allard, \& Olinto 2010~\cite{Kotera2010} lower& 0.005  & 4.1  \\
{\it Models constrained by Fermi cascade bound:}&  &  \\
~~~~~~Ahlers {\it et al.} 2010~\cite{Ahlers2010}& 0.09  & 20.7 \\
{\it Waxman-Bahcall (WB) fluxes}: & & \\
~~~~~~WB 1999, evolved sources~\cite{WB}~~ & 1.5 & 76 \\
~~~~~~WB 1999, standard~\cite{WB} &  0.5 &   27  \\ 
  \end{tabular}
 \end{footnotesize}
\end{table}

%

In Table~\ref{nurates} we give expected neutrino event totals from a wide
range of currently allowed cosmogenic neutrino models for ARA in three
years of operation, compared to recent published expectations for the best current limits
to date, from the ANITA-II flight~\cite{ANITA2}. It is evident that ARA-37 will
extend in sensitivity above ANITA-2's sensitivity by factors of
two orders of magnitude or more.
For strong-source-evolution and
baseline models,  ARA-37 detects between of order 50 to over 200 events 
in three years of operation, 
enough to establish the
basic characteristics of the energy spectrum and source arrival directions.

There are also recent cosmogenic neutrino flux estimates which 
compute neutrino fluxes subject to constraints
from the Fermi diffuse gamma-ray background~\cite{Ahlers2010}, and which
include a heavier nuclear composition ({\it e.g.}, an admixture of iron) 
for the UHECRs~\cite{Ave05,Stanev07,Kotera2010}.
Over a 3-year timescale all of these models are detectable, but in some cases 
only marginally, and up to five years will be necessary to establish the flux.
Over the planned instrument life of a decade or more, ARA-37 will thus be
able to not only establish the flux levels for all of even the most conservative models,
but to begin measurements of their energy spectral dependence as well.

		\subsection{Resolution}

Although not directly important for detection of neutrinos, the resolution of
both the distance and angles to the neutrino interaction vertex, as well as
the ability to reconstruct coarse neutrino incident directions on the sky, are
important characteristics of our detector, and we have studied them in detail.
This is especially important for our current realization of ARA-37, since
the wider spacing will lead to very few multi-station coincident events,
and thus each station must function as a stand-alone neutrino detector in
both shower energy estimation and neutrino direction angular resolution.

To make these measurements, we have 16 antennas per station, and thus
16 waveform amplitudes and phases, as well as the frequency spectral components 
of the coherently-summed waveform which can be estimated to good precision
once the arrival direction is fitted. From the Vpol and Hpol data we also
fit the plane of polarization, and with precise
timing we can measure the radius of curvature of the arriving wavefront.

Our measurement of the distance to the neutrino vertex is accomplished by
the estimates of the wavefront curvature. 
This may be thought of as measuring the residuals when
fitting the arrival times to a plane wave. For the angular measurements, 
the antenna array is analyzed as a correlation interferometer, and precise
timing differences between the arrival times of the Askaryan radio impulse are
determined for all of the $N(N-1)/2$ pairs of $N$ antennas.

Complementing the precise timing measurements, we can also operate our cluster
array as a radio intensity gradiometer and polarimeter. The gradiometric function comes
through amplitude calibration of the received impulse, and the polarimetric
information comes from ratios of the calibrated amplitudes of the Vpol and Hpol
antennas.

\begin{figure*}[t!]
 \centerline{\includegraphics[width=2.2in]{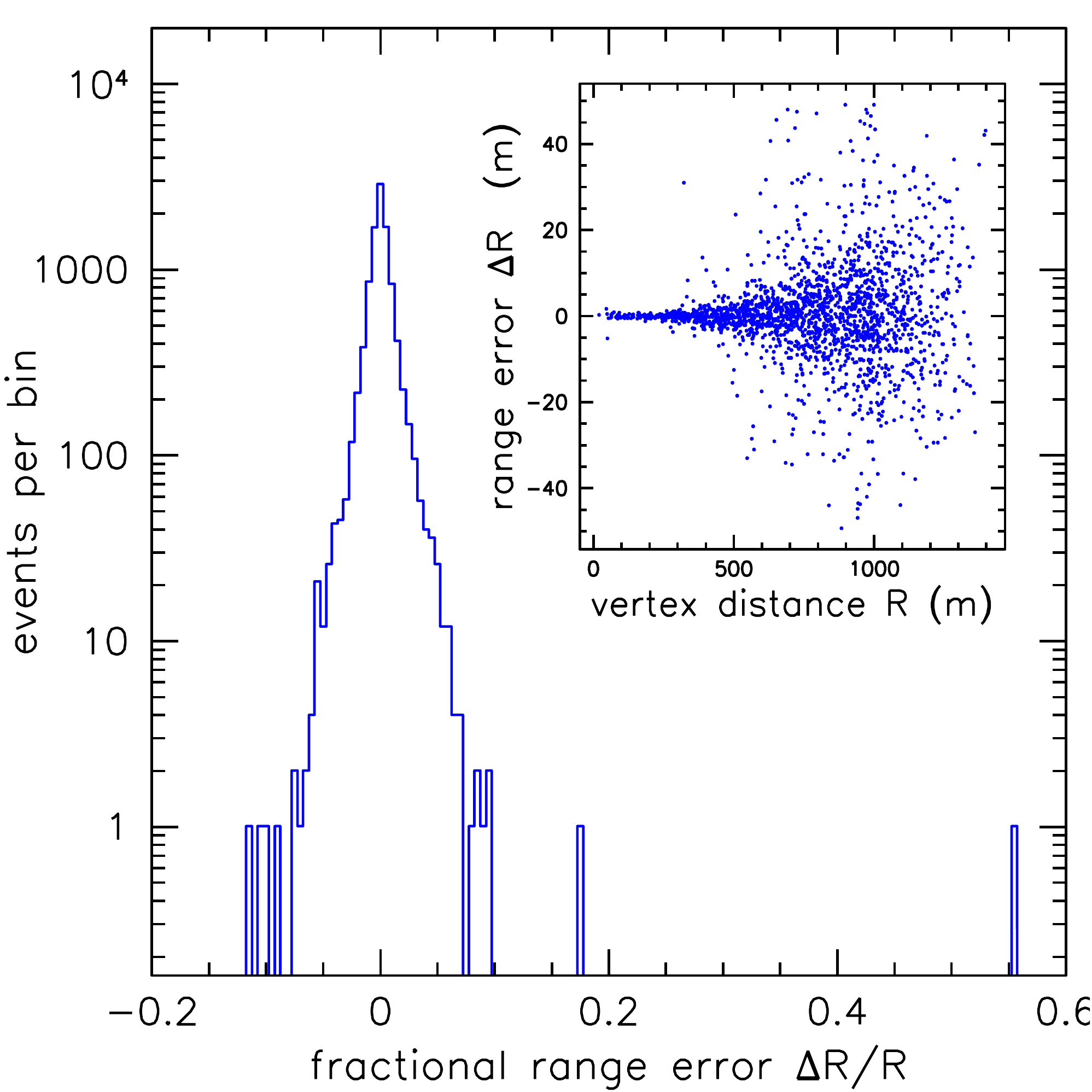}~~\includegraphics[width=2.2in]{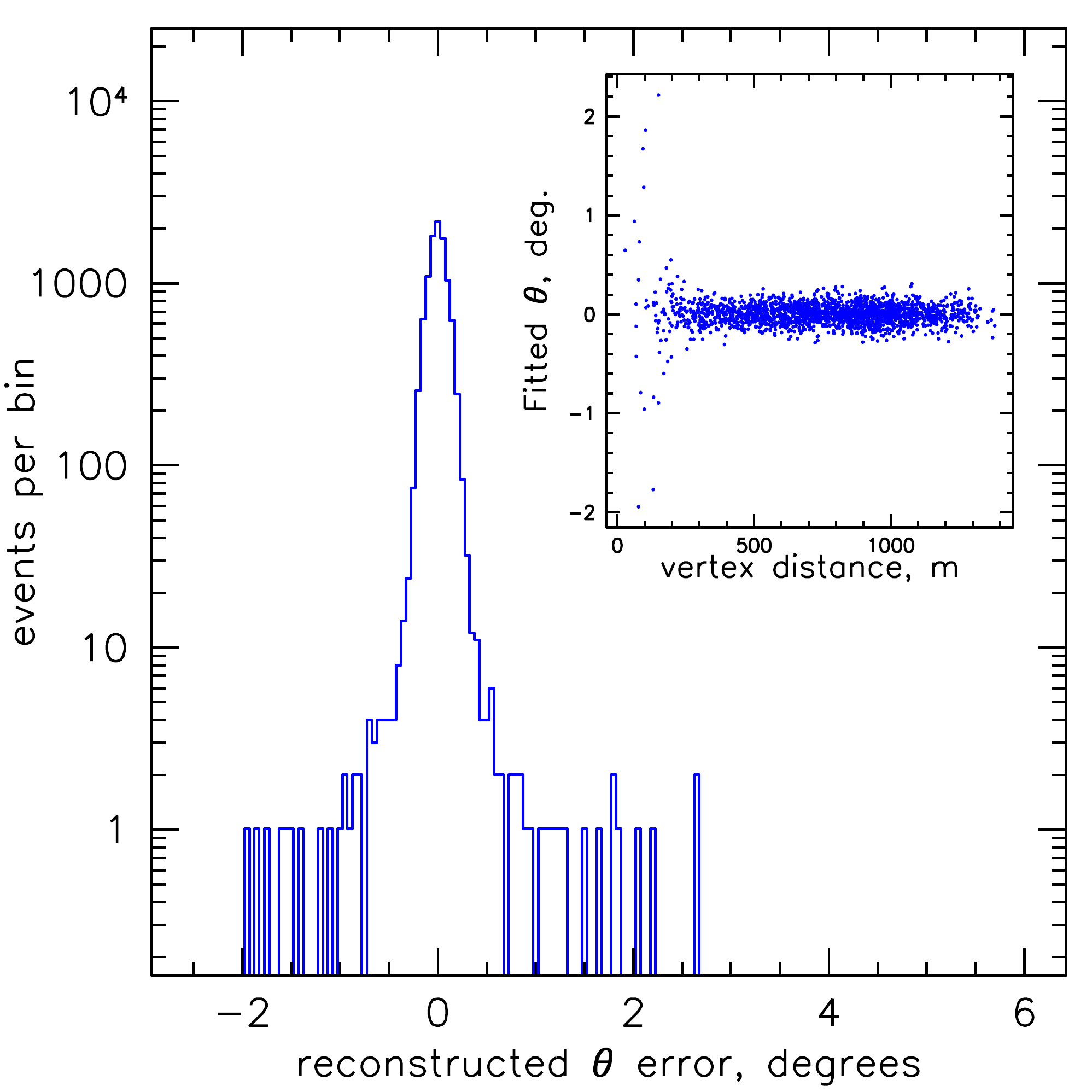}~~\includegraphics[width=2.3in]{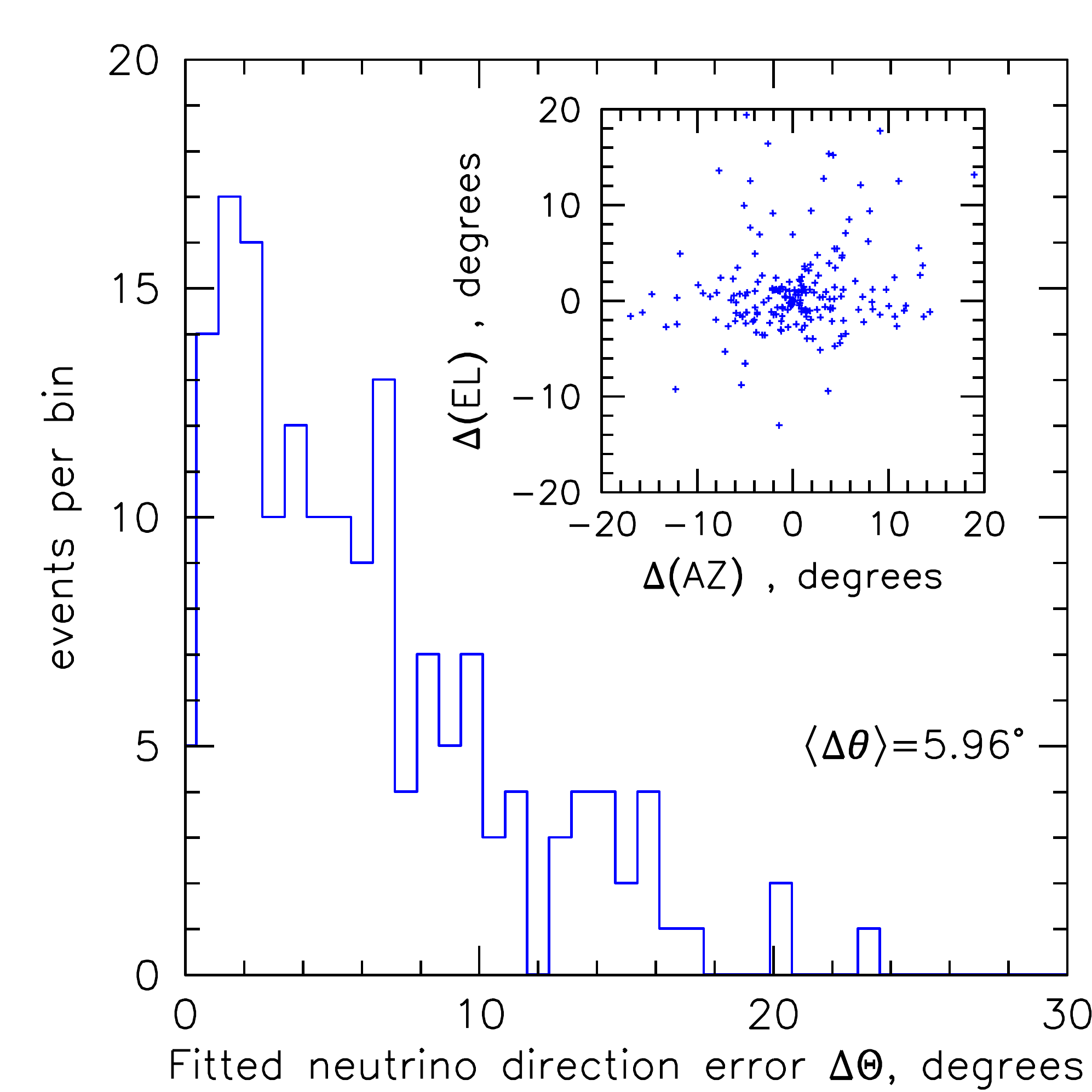}}
\begin{small}
\caption{ Left: Distribution of fractional range errors in single-antenna-cluster reconstruction of neutrino interaction vertex. 
Left inset: the distribution of reconstructed vertex range errors vs. range. Middle: Similar to left plot
for reconstructed zenith angle of vertex relative to antenna cluster. Middle inset: distribution of 
reconstructed zenith angle vs. range to vertex. Right: Distribution of polar angle errors for
full reconstruction of the incoming neutrino direction, using vertex reconstruction, amplitude, and 
polarization information. Right inset: the 2-D distribution of reconstructed directions relative to
true neutrino direction.
    \label{VertexSim}
  }
\end{small}
\end{figure*}

All of these estimates are done in offline reconstruction routines. They are
not necessary for the triggering of the array to record potential neutrino events,
but they do make maximal use of the recorded information in the waveforms 
and arrival times of the events.

			\subsubsection{Vertex Resolution}

The critical parameter for vertex location is the intra-cluster timing precision.
For this we have used actual measurements made with ANITA data, to which our collaboration
has access. The ANITA payload, which uses waveform digitizers that are 
comparable to our planned digitizers, has demonstrated timing resolution as
good as 30~ps rms for waveforms registered at the $4\sigma$-level detection
threshold of ANITA. These timing precisions come about from extensive in-flight
calibration using ground-based impulse generators, and have proven robust
in the ANITA analysis~\cite{ANITAinst}. For our simulations we have derated these
values by a factor of 3.3 to account for our more limited radio bandwidth, 
the slower sampling rate we expect to use, and for
possibly unknown systematics in our
calibration.

Fig.~\ref{VertexSim}(left,middle) shows the results of these simulations for both the
range and pointing resolution to the vertex.
The latter values are important
for determining whether an event originates under the ice or from
above the ice, and the former values, combined with our knowledge of the
ice attenuation, will bear directly on our ability to perform
calorimetry on the neutrino shower.

			\subsubsection{Incident neutrino direction resolution}
Estimating the neutrino vertex location in three dimensions does not immediately
determine the arrival direction of the neutrino itself, and it may appear that
the tight constraints of the cluster geometry would preclude determination
of the neutrino direction, since this is normally done by imaging of the
Cherenkov cone, at least in ring-imaging Cherenkov detectors. However, the
richness of the radio wave information content does in fact allow for first
order Cherenkov cone determination, using both the amplitude gradient of 
the cone, and the polarization vector, which lies in the plane containing the
Poynting vector of the radiation and the shower momentum vector. Thus
once the Poynting vector is determined via the vertex reconstruction, the
plane of polarization combined with the local gradient in the cone amplitude
is sufficient to constrain the neutrino direction on the sky.

To study the neutrino direction resolution, we have included first-order
reconstruction algorithms in our Monte Carlo neutrino simulation. These
do not yet perform a maximum likelihood minimization which would be very
appropriate for this complex problem, but instead they perform
a $\Xi^2$-grid search over variational parameters once the event has
been detected in the simulation, which includes appropriate thermal noise
backgrounds. These simulations are computationally intensive, and involve
full 3D ray-tracing of the radio propagation through the ice for each tested
grid-point. Fig.~\ref{VertexSim}(right)
shows the results for about 80 detected events in a simulation run
at $E_{\nu}=3 \times 10^{18}$~eV. The reconstruction for these events
was about 80\% efficient (that is, 20\% of the events failed to reconstruct),
but the reconstructed zenith angle distribution shown has a very acceptable standard error
of $\sim6^{\circ}$, the simplicity of our reconstruction code notwithstanding. 

			\subsubsection{Energy Resolution}
Once the event geometry has been reconstructed, the shower energy is determined
via standard parametric equations~\cite{ZHS,Alv97,FORTE}. 
For example, a parameterization due to
Lehtinen~\cite{FORTE} gives the shower
energy as a function of radio frequency $E_{sh}(f)$ as:
\begin{equation}
|E_{sh}(f;R,\theta)| =\sqrt{2\pi}R^{-1}\mu\mu_0~Q~L~f\sin\theta ~e^{-(kL)^2(\cos\theta-1/n)^2/2}
\label{eq:cher}
\end{equation}
where $R$ is the shower distance, $Q$ and $L$ are standard 
shower parameters relating to the excess charge
and shower length, $\theta$ is the observer's polar angle with respect to the
shower (and neutrino) direction, $k=2\pi/\lambda$ is the wavenumber, $n$ the index of refraction
of the ice; for typical dielectrics $\mu=1$, and $\mu_0= 4\pi \times 10^{-7}$ is
the permeability of free space. 

Using standard variational
analysis the fractional shower energy ($E_{sh}$) variation with range error $\Delta R$ is 
$|\Delta E_{sh}/E_{sh}|_{R} = e^{-\alpha R}(\Delta R/R)$ for small ice attenuation 
coefficient $\alpha = L_{\alpha}^{-1}$.
The variation with uncertainty in $\alpha$ is
$|\Delta E_{sh}/E_{sh}|_{\alpha} = Re^{-\alpha R}(\Delta\alpha)$ 
which is generally negligible since $\Delta\alpha$ is of order $10^{-3}$ at most.
The corresponding variation with neutrino angular error $\Delta\theta$ is 
$|\Delta E_{sh}/E_{sh}|_{\theta} \simeq \cos\theta_C \Delta\theta$,
where we evaluate the variation near the Cherenkov angle $\theta_C$.
Finally an estimate of the neutrino energy from shower energy must account
for the large variance in the Bjorken-$y$-distribution, 
defined by $E_{\nu}\simeq y^{-1}E_{sh}$ for charged-current $\nu_{\mu},\nu_{\tau}$
and all neutral-current events; electron neutrino events will be less
affected by this, so our estimate is conservative.
At neutrino energies of $10^{17-19}$~eV, numeric evaluations give 
$\langle y \rangle \simeq 0.22, ~\Delta y/y \simeq 1$,
thus $|\Delta E_{\nu}/E_{\nu}|_{y} = \Delta y/\langle y \rangle \simeq 1$.
Assuming these errors are uncorrelated, and using $\Delta R/R \sim 0.02$
with a mean $R\sim 1$~km, and $cos\theta_C~ \delta\theta = 0.06$, the
root-sum-squared error is dominated by the Bjorken-$y$ uncertainty, giving 
$|\Delta E_{\nu}/E_{\nu}|_{total} \sim 1$
for $E_{\nu} = 3 \times 10^{18}$~eV. 
This resolution will also be comparable for lower neutrino energies in
the GZK neutrino spectral range. The $y$-dominated uncertainty
is generic for UHE neutrino experiments, but this energy resolution is wholly adequate for
the first-order science goals of the ARA instrument.

		\subsection{Comparison to Existing Instruments}

\begin{figure}[tb!]
\includegraphics[width=3.5in]{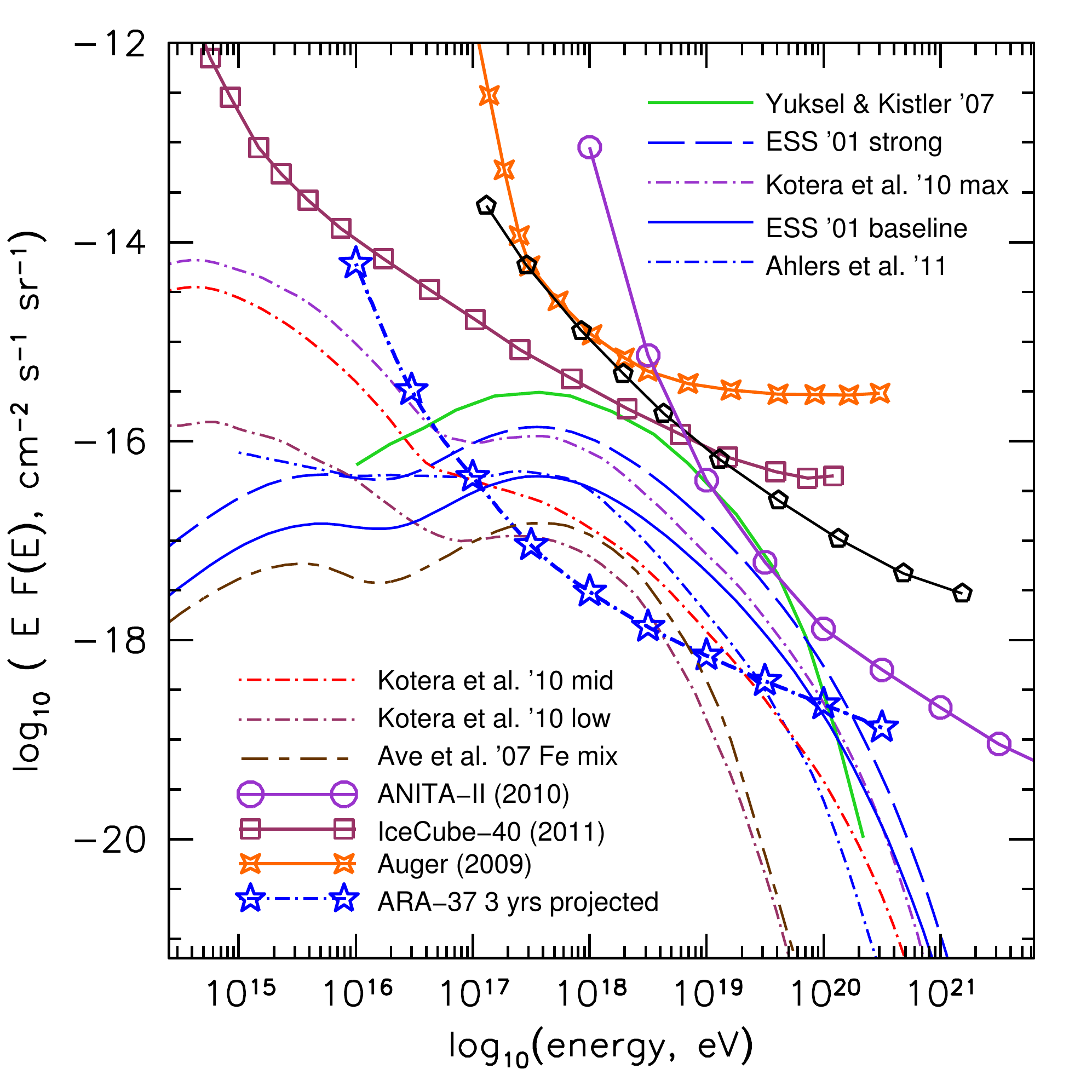}
\begin{small}
\caption{ Compilation of sensitivity estimates from existing instruments,
published limits, and a range of GZK neutrino models, along with the
expected 3 year ARA sensitivity.
  }
\label{NewGZK11}
\end{small}
\end{figure}

Fig.~\ref{NewGZK11} provides a comprehensive graphic summary of the comparison of our
estimated ARA sensitivity to estimates for several operating experiments,
along with 2006 limits from the ARA forerunner experiment RICE~\cite{RICE06}. We
have already noted the comparison of ARA to the published ANITA limits;
here we use projections for ANITA's reach after three flights, along
with similar projections for IceCube and the Auger Observatory. GZK neutrino
models are also included from a wide range of 
estimates~\cite{Kal02,Kal02a,PJ96,Engel01,WB,Ave05,Aramo05}, including
the pure-Iron UHECR composition model noted above. 

ARA improves over any other current instrument by an
order of magnitude within 3 years of operation, filling in an important
gap in sensitivity in the heart of the cosmogenic neutrino spectral energy
region. IceCube has excellent sensitivity to lower energies, up to the
10~PeV level, and
ANITA has unmatched sensitivity at the higher energies, above 10~Eev. 
The Auger Observatory, while probing a similar energy range as ARA,
does not have as high a neutrino sensitivity as it is primarily a UHECR
instrument. ARA will complement these other instruments by making high sensitivity
observations in the 0.1-10~EeV energy range, matching the peak of the
expected cosmogenic neutrino fluxes.

\section{Conclusions}

We have described the design and initial performance of a new
ultra-high energy neutrino detector at the South Pole, the
16-antenna, self-triggering ARA-testbed, which is a high-fidelity prototype for future
ARA detector stations. Our initial operation extending well into
the the extreme thermal environment of the austral winter indicates that
radio-frequency interference is infrequent and has only a slight
impact on operation for our testbed detector, which is closest of any
future ARA stations to the primary sources of interference at
the South Pole station. Other than brief periods of sporadic 
interference, the baseline radio noise levels are dominated by
the pure thermal noise floor of the ambient ice, and the
thermal noise does not appear to be correlated to wind velocity.
We have demonstrate the ability to maintain impulse trigger sensitivity
at a level close to the thermal noise.
We have demonstrated RF impulse propagation of more than 3~km slant
range through the South Pole ice without significant loss of
signal coherence. We have demonstrated inter-antenna pulse timing
precision of order 100~ps, implying angular resolutions which
are more than adequate for neutrino vertex reconstruction.
We have presented simulations using characteristics projected
from our measurements which give high confidence that our 
completed phase-I array, ARA-37, will achieve its goal of
a robust detection of cosmogenic neutrinos, and will lay a clear
foundation for an observatory-class instrument.

\section*{Acknowledgments}
We thank the National Science Foundation for their generous support through Grant NSF OPP-1002483.
We are very grateful to Raytheon Polar Services Corporation for their excellent field support at Amundsen-Scott
Station, and the IceCube drillers and specialists who helped with our
installation.

\appendix
\section{ARA Autonomous Renewable Power Stations (AARPS)}

As ARA moves farther from the station, the transition from station power to autonomous power sources will become increasingly important.  The planned ARA footprint calls for three ARA stations to be powered from a single node, requiring about 300W from that node.

A variety of power sources were reviewed during 2010 including photovoltaic (PV) arrays, wind turbines, diesel generators, fuel cells, and Stirling engine generators.  The first three remain in consideration with the renewable sources, PV and wind, to be attempted first.  PV is well-known to be rugged and efficacious on the Antarctic plateau, so the 2010-2011 season was instead used to study wind power options for the Antarctic plateau.

The objectives in the first season of deployment were to:
\begin{itemize}
\item Test wind turbine candidates for survivability and for power production.  While several larger and several smaller turbines have been deployed on the plateau, there have been no known studies in the intermediate
power range that ARA requires.
\item Measure wind speed as a function of height from the surface.  This profile, determined by the turbulence produced by the surface upwind of the turbine, will dictate the required tower height.
\item Understand tower erection techniques suitable for the South Pole environment.
\item Fully instrument the system with sensors for temperature at various locations, power from turbines, wind speeds, etc. connected to a custom System Health Monitor (SHM).  Those data are transmitted to the ICL.
\item Test a Power Instrument Box (PIB). The PIBs deployed are thermally insulated but do not have rf shielding; shielding is planned for future PIBs.
\item Test for electrical noise in the 200MHz to 800MHz band which could result from the turbine itself, the triboelectric effect on the tower, or emission from instrumentation.
\item Test low-power Zigbee communications.
\end{itemize}
The ARA effort to develop autonomous power stations at Pole began in the 2010-2011 season with the deployment of 3 small wind turbines, each with a rated power output of slightly over a kw:

\begin{figure}[htpb]
\centerline{\includegraphics[width=8cm,angle=0]{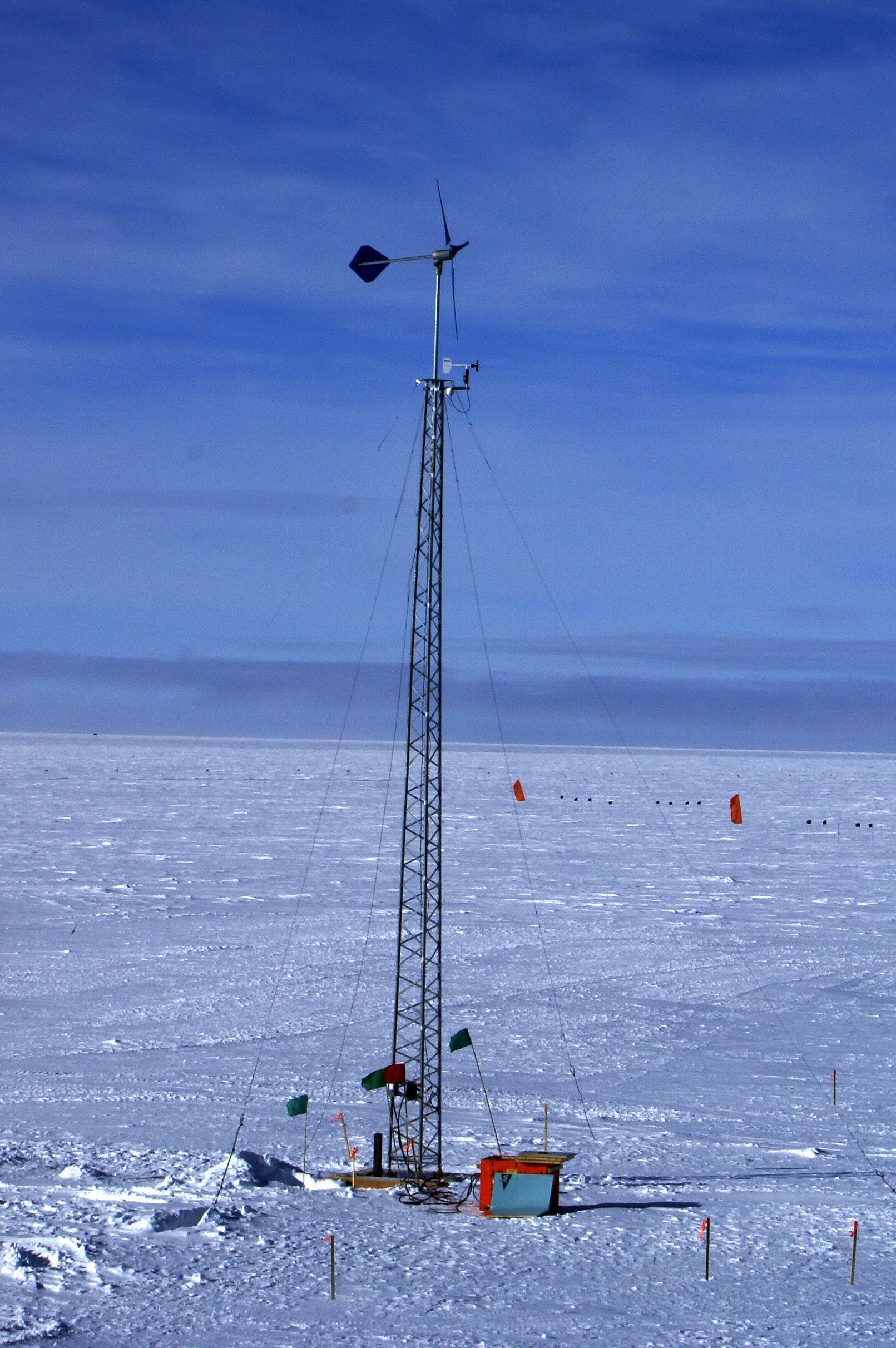}}\caption{Raum model turbine, installed at South Pole in January, 2011.}
\label{fig:Raum.jpg}
\end{figure}
At Site 1, a Raum 1kw turbine (Fig. \ref{fig:Raum.jpg})
was installed on a 50-ft lattice tower.  It was raised on a tilt-up tower with a Pisten Bully and a fixed 10’ gin pole.  Four guy wires are anchored by timbers placed as dead men in trenches in the ice.  A single anemometer is mounted just below the turbine.  

\begin{figure}[htpb]
\centerline{\includegraphics[width=8cm,angle=0]{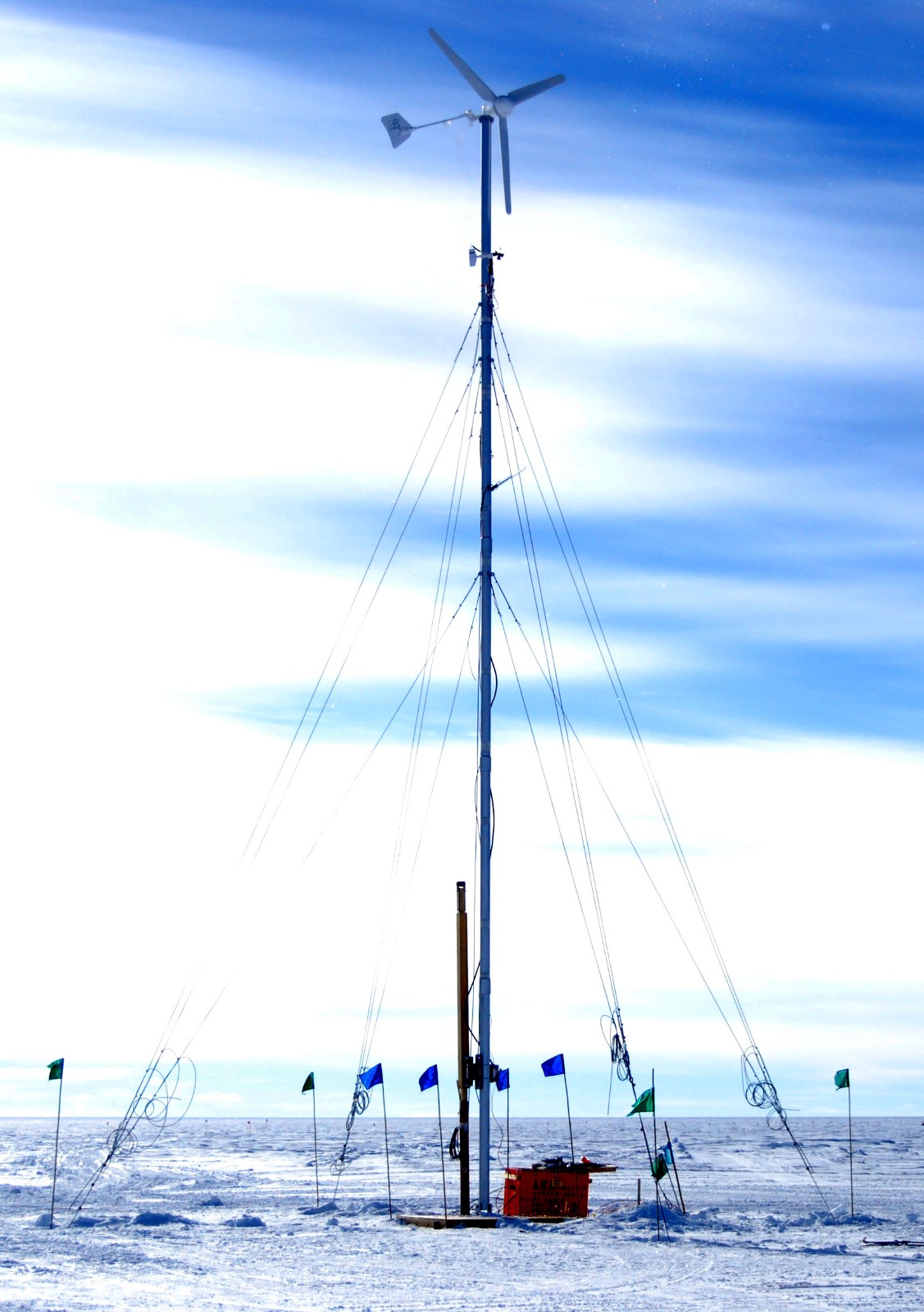}}\caption{Hummer model turbine, installed at South Pole in January, 2011.}
\label{fig:Hummer.jpg}
\end{figure}
At Site 2, a Hummer kw turbine (Fig. \ref{fig:Hummer.jpg}) was deployed on a 50’ monopole.  The greater weight of the turbine and tower required the use of a Caterpillar 853 to raise the turbine.  Three sets of four guy wires are anchored by dead men.  A single anemometer is mounted just below the turbine.  A Yagi antenna is mounted below the anemometer and pointed toward the ICL.

\begin{figure}[htpb]
\centerline{\includegraphics[width=8cm,angle=0]{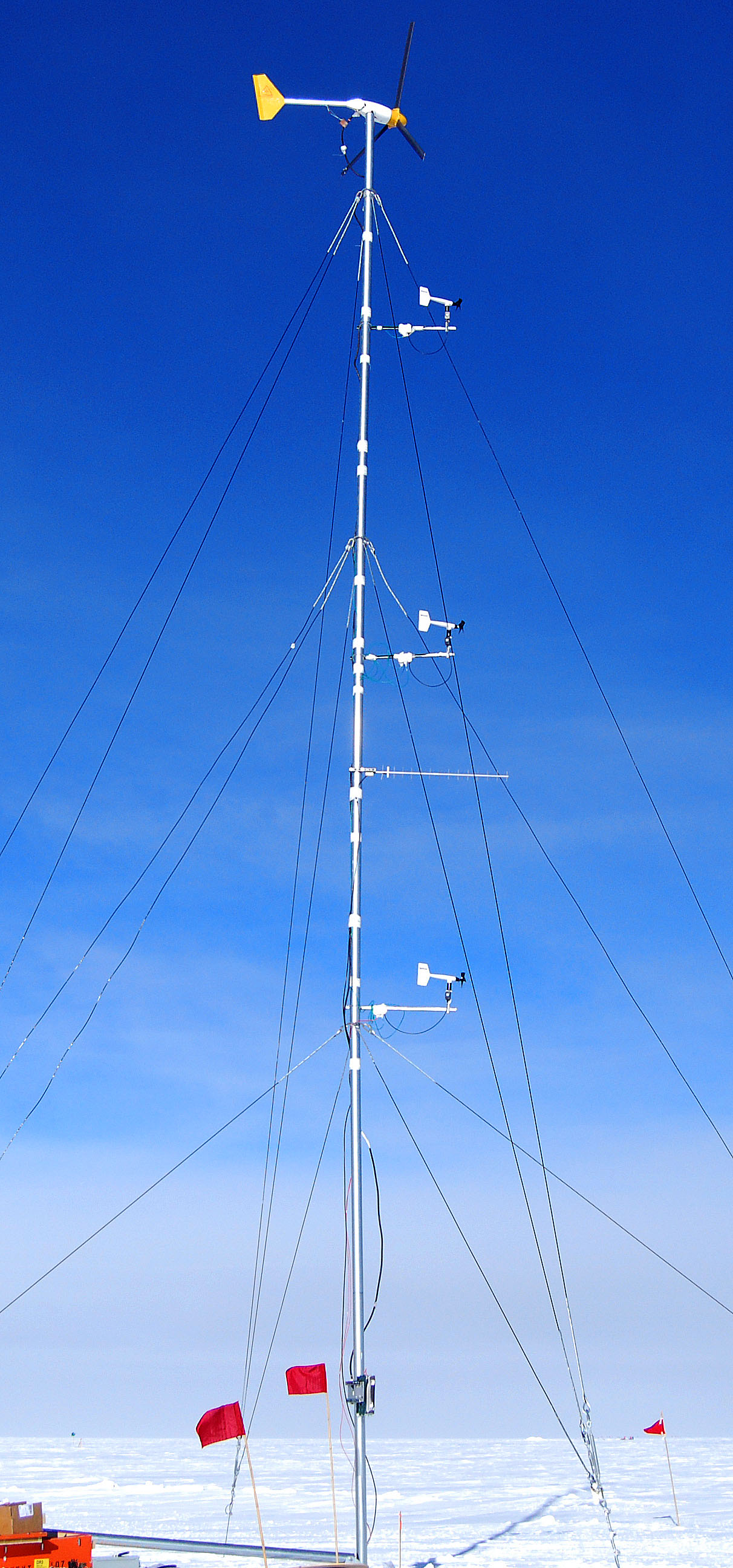}}\caption{Bergey model turbine, installed at South Pole in January, 2011.}
\label{fig:Bergey.jpg}
\end{figure}
At Site 3, a Bergey XL.1 turbine (Fig. \ref{fig:Bergey.jpg}) was raised on a 60’ monopole and integral 20’ gin pole using a Pisten Bully.  Three sets of four guys were installed and anchored by dead men.  In this location only, Kevlar guys are used rather than steel cables to reduce static build-up.  A Yagi antenna was mounted on this tower as well.

\begin{figure}[htpb]
\centerline{\includegraphics[width=8cm,angle=0]{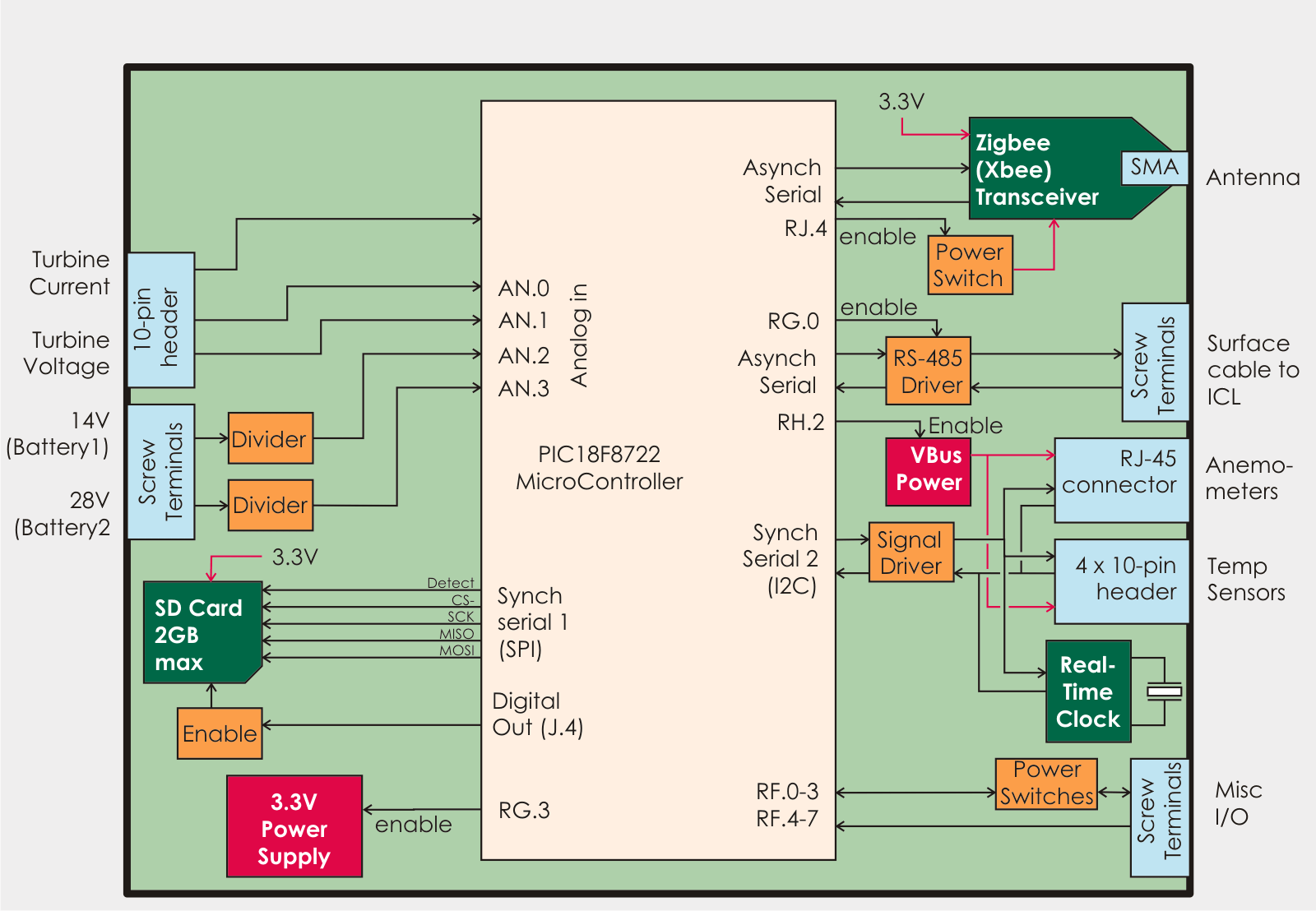}}\caption{Schematic of System Health Monitor (SHM) installed at South Pole in January, 2011.}
\label{fig:SHM.png}
\end{figure}
Each turbine was outfitted nearly identically, including the following common characteristics. i) A brake switch and high-power dump resistors are mounted on each tower. ii) An SHM (Fig. \ref{fig:SHM.png}) is located inside each PIB to collect metrology and power data and store and/or transmit the data.
The power is routed to dump resistors inside and outside the box 
\begin{figure}[htpb]
\centerline{\includegraphics[width=8cm,angle=0]{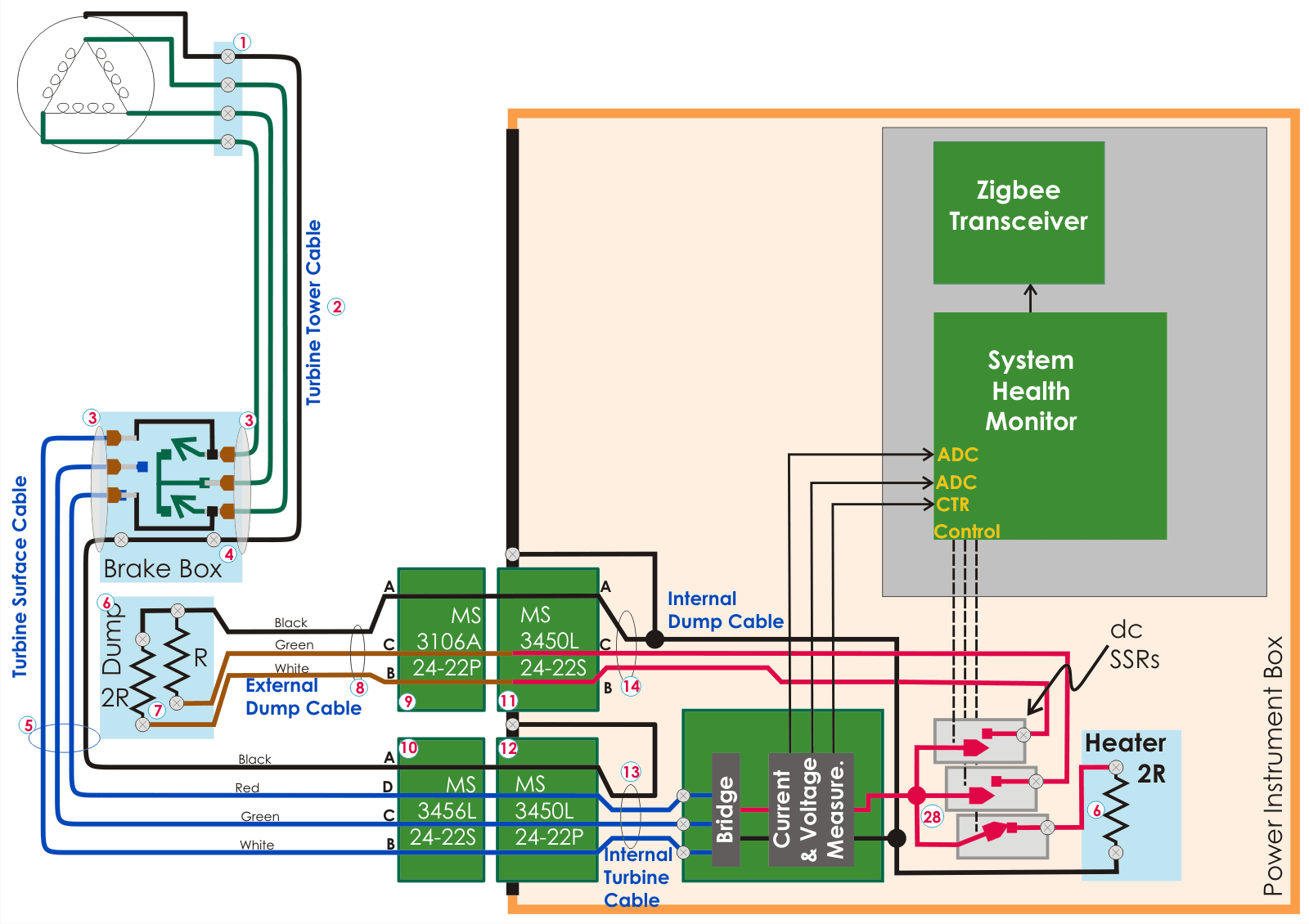}}\caption{System Health Monitor system installed at South Pole in January, 2011.}
\label{fig:PS.png}
\end{figure}
(Fig. \ref{fig:PS.png}), controlled by the SHM to maintain a warmed environment inside the PIB.
iii) The SHM provides three destinations for the data:  a 2GB SD card for local storage, a Zigbee rf link (operated only for short-term testing until approval is obtained), and a hard surface link to the ICL.
A Power Board containing a three-phase rectifier and measurement of voltage, current and rotational rate.

\subsection{Observations}

\subsubsection{Power Output}

\begin{figure}[htpb]
\centerline{\includegraphics[width=8cm,angle=0]{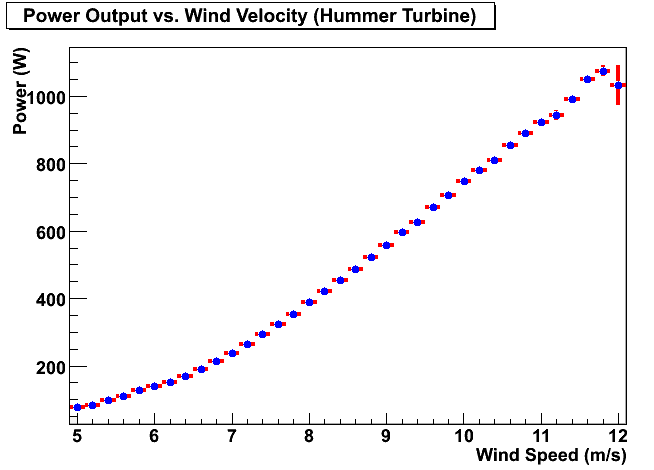}}\caption{Plot of turbine output power vs. velocity, illustrating expected power law relationship. A fit gives P(velocity)$\propto v^{2.8}$, roughly consistent with the expected cubic dependence.}
\label{fig:HummerPower.png}
\end{figure}
Representative data from one turbine are plotted as a function of wind speed in Fig. \ref{fig:HummerPower.png}.  Very roughly we see the expected relationship of the cube of the wind speed.

\subsubsection{Wind Speed as a function of height}
\begin{figure}[htpb]
\centerline{\includegraphics[width=8cm,angle=0]{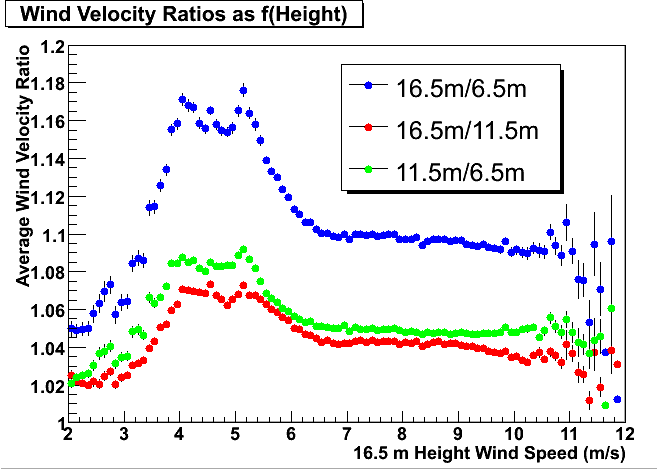}}\caption{
Ratios of wind speeds at various height combinations, illustrating the increase in velocity with elevation above snow surface.}
\label{fig:BergeyMetData.png}
\end{figure}
The data collected at three heights on the Bergey tower are shown in Figure \ref{fig:BergeyMetData.png}. It shows the height relationship to be relatively mild, but nevertheless illustrates the benefit of taller towers, especially at lower wind speed.

\subsubsection{Modeling Turbine Performance}
\begin{figure}[htpb]
\centerline{\includegraphics[width=8cm,angle=0]{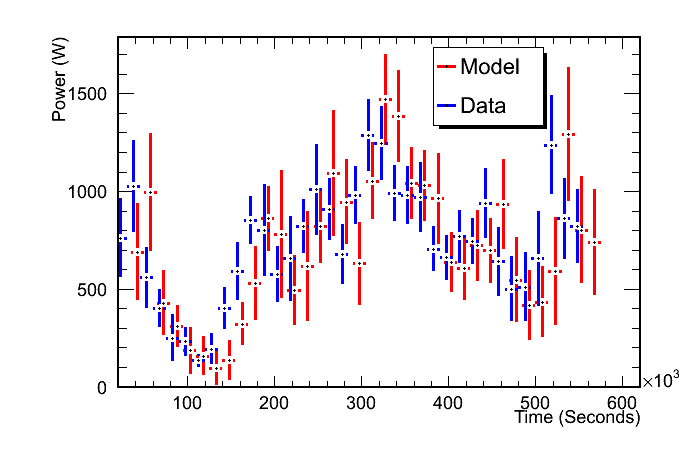}}\caption{
Comparison of modeled vs. measured wind turbine performance; data taken from Jan. 26--Feb. 1.}
\label{fig:RHummerPowerModel.png}
\end{figure}

To project the future duty cycle a photovoltaic + wind-based system will offer, a detailed turbine simulation has been developed. This 
calculation incorporates turbine-specific manufacturers' data, in addition to our own 2011 
measurements of output power vs. wind speed (Fig. \ref{fig:HummerPower.png})
and wind velocity profile dependence as a function of above-surface elevation (Fig. \ref{fig:BergeyMetData.png}). For the particularly high-wind
conditions at South Pole during the last week of January, 2011, we find agreement between our predicted vs. actual performance at the 5--10\% level,
as shown in Figure \ref{fig:RHummerPowerModel.png}. Without the aid of batteries, based on our model, and folding in measured wind speed data from the South Pole over the last half-decade, we expect overall power needs to be satisfied by a photovoltaic+wind-turbine system at least 92\% of the time. Buffering power with 48-hour batteries should improve the duty cycle to $>$97\%.


\subsubsection{Data Collection}

The SHMs have operated reliably, collecting data from the sensors and sending data to the ICL.  The array of temperature and metrology sensors is operating reliably.  Two of the three custom power sensor boards continue to operate reliably; the one failure may be the Power Board or the cabling and will be investigated during the 2011-12 austral summer.
The SD card operation was verified by retrieving the card after several weeks and downloading to a laptop.  A simple 4” whip antenna was able to receive low-power Zigbee data at the ICL from all three sites; a mesh network is planned for next season.  Data are being received by surface cable using the RS-485 protocol from two sites.  
Received data are being shipped to the north at regular intervals.
The SHM that was deployed consumes over 2 watts, but the design cycle and firmware development now underway are bringing that value down substantially.
A custom acquisition and transmission system for the RM Young anemometers also operates at very low power.  The multiple anemometers, temperature sensors and planned vibration sensor can be daisy-chained to simplify connection.

\subsubsection{Power Instrumentation Box}

These were buried in the ice to provide wind protection.  Although we have some modest temperature control, we do not yet have sufficient information to determine how much energy is required to maintain the boxes at 0C.  Structurally the PIBs worked well with the proviso that the cable penetrations need modification to facilitate access.

\subsubsection{Electrical noise from AARP sites}
As mentioned elsewhere in this document,
attempts to “see” noise from the turbine sites with the test bed have been negative.

\end{document}